\documentclass[modern]{aastex62}

\received{\today}
\revised{\today}
\accepted{\today}
\submitjournal{ApJ}

\usepackage{savesym}
\usepackage{amsmath,amssymb}

\usepackage{graphicx}
\usepackage{hyperref}
\usepackage[version=4]{mhchem}
\usepackage{url}
\usepackage{float}

\usepackage{amsmath,amstext,amssymb}

\usepackage[T1]{fontenc}
\usepackage{tikz}
\usetikzlibrary{arrows}
\usetikzlibrary{positioning}
\usetikzlibrary{shapes,decorations}
\usepackage[figure,figure*]{hypcap}

\shorttitle{Acetylene as a chemical tracer for impacts}
\shortauthors{Rimmer et al.}

\begin{document}

\title{Identifiable Acetylene Features Predicted for Young Earth-like Exoplanets with Reducing Atmospheres undergoing Heavy Bombardment}

\correspondingauthor{P.~B. Rimmer}
\email{pbr27@cam.ac.uk}

\author{P.~B. Rimmer}
\affiliation{University of Cambridge, Department of Earth Sciences Downing St, Cambridge CB2 3EQ}
\affiliation{University of Cambridge, Cavendish Astrophysics, JJ Thomson Ave, Cambridge CB3 0HE}
\affiliation{MRC Laboratory of Molecular Biology, Francis Crick Ave, Cambridge CB2 OQH}

\author{M. Ferus}
\affiliation{J. Heyrovsky Institute of Physical Chemistry, Czech Academy of Sciences, Dolej\v{s}kova 2155/3, 182 23 Prague, Czech Republic}

\author{I.~P. Waldmann}
\affiliation{Department of Physics \& Astronomy, University College London, Gower Street, London, WC1E 6BT}

\author{A. Kn\'{i}\v{z}ek}
\affiliation{J. Heyrovsky Institute of Physical Chemistry, Czech Academy of Sciences, Dolej\v{s}kova 2155/3, 182 23 Prague, Czech Republic}
\affiliation{Department of Physical and Macromolecular Chemistry, Faculty of Science, Charles University, Albertov 2030, 128 43 Prague, Czech Republic}

\author{D. Kalvaitis}
\affiliation{University of Cambridge, Department of Chemistry, Lensfield Road, Cambridge CB2 1EW}

\author{O. Ivanek}
\affiliation{J. Heyrovsky Institute of Physical Chemistry, Czech Academy of Sciences, Dolej\v{s}kova 2155/3, 182 23 Prague, Czech Republic}

\author{P. Kubel\'{i}k}
\affiliation{J. Heyrovsky Institute of Physical Chemistry, Czech Academy of Sciences, Dolej\v{s}kova 2155/3, 182 23 Prague, Czech Republic}
\affiliation{Institute of Physics, Czech Academy of Sciences, Na Slovance 1999/2, 182 21 Prague, Czech Republic}

\author{S.~N. Yurchenko}
\affiliation{Department of Physics \& Astronomy, University College London, Gower Street, London, WC1E 6BT}

\author{T. Burian}
\affiliation{J. Heyrovsky Institute of Physical Chemistry, Czech Academy of Sciences, Dolej\v{s}kova 2155/3, 182 23 Prague, Czech Republic}
\affiliation{Institute of Physics, Czech Academy of Sciences, Na Slovance 1999/2, 182 21 Prague, Czech Republic}
\affiliation{Institute of Plasma Physics, Czech Academy of Sciences, Za Slovankou 1782/3, 182 00 Prague Czech Republic}

\author{J. Dost\'{a}l}
\affiliation{Institute of Physics, Czech Academy of Sciences, Na Slovance 1999/2, 182 21 Prague, Czech Republic}
\affiliation{Institute of Plasma Physics, Czech Academy of Sciences, Za Slovankou 1782/3, 182 00 Prague Czech Republic}

\author{L. Juha}
\affiliation{Institute of Physics, Czech Academy of Sciences, Na Slovance 1999/2, 182 21 Prague, Czech Republic}
\affiliation{Institute of Plasma Physics, Czech Academy of Sciences, Za Slovankou 1782/3, 182 00 Prague Czech Republic}

\author{R. Dud\v{z}\'{a}k}
\affiliation{Institute of Plasma Physics, Czech Academy of Sciences, Za Slovankou 1782/3, 182 00 Prague Czech Republic}

\author{M. Kr\r{u}s}
\affiliation{Institute of Plasma Physics, Czech Academy of Sciences, Za Slovankou 1782/3, 182 00 Prague Czech Republic}

\author{J. Tennyson}
\affiliation{Department of Physics \& Astronomy, University College London, Gower Street, London, WC1E 6BT}

\author{S. Civi\v{s}}
\affiliation{J. Heyrovsky Institute of Physical Chemistry, Czech Academy of Sciences, Dolej\v{s}kova 2155/3, 182 23 Prague, Czech Republic}

\author{A.T. Archibald}
\affiliation{University of Cambridge, Department of Chemistry, Lensfield Road, Cambridge CB2 1EW}
\affiliation{National Centre for Atmospheric Science, University of Cambridge, Cambridge CB2 1EW}

\author{A. Granville-Willett}
\affiliation{University of Cambridge, Department of Chemistry, Lensfield Road, Cambridge CB2 1EW}

\begin{abstract}
The chemical environments of young planets are assumed to be largely influenced by impacts of bodies lingering on unstable trajectories after the dissolution of the protoplanetary disk. We explore the chemical consequences of impacts within the context of reducing planetary atmospheres dominated by carbon monoxide, methane and molecular nitrogen. A terawatt high-power laser was selected in order to simulate the airglow plasma and blast wave surrounding the impactor. The chemical results of these experiments are then applied to a theoretical atmospheric model. The impact simulation results in substantial volume mixing ratios within the reactor of 5\% hydrogen cyanide (HCN), 8\% acetylene (C$_2$H$_2$), 5\% cyanoacetylene (HC$_3$N) and 1\% ammonia (NH$_3$). These yields are combined with estimated impact rates for the Early Earth to predict surface boundary conditions for an atmospheric model. We show that impacts might have served as sources of energy that would have led to steady-state surface quantities of 0.4\% \ce{C_2H_2}, 400 ppm HCN and 40 ppm \ce{NH_3}. We provide simulated transit spectra for an Earth-like exoplanet with this reducing atmosphere during and shortly after eras of intense impacts. We predict that acetylene is as observable as other molecular features on exoplanets with reducing atmospheres that have recently gone through their own `Heavy Bombardments', with prominent features at 3.05 $\mu$m and 10.5 $\mu$m.
\end{abstract}


\section{Introduction} \label{sec:intro}

The chemical evolution of the Earth and chemical origin of life are two of the most fascinating and most controversial questions asked in contemporary science. Direct geological records revealing the chemical evolution of Earth's atmosphere during the Hadean eon (> 4 Gya) of the early Earth are sporadic and indirect. The general question of when life originated is a hotly contested point in the community, with estimates ranging from 3.5 Gya \citep{Schopf1987}, to 3.8 Gya or even 4.1 Gya \citep{Mojzsis1996,Bell2015,Dodd2017}.

From a combination of planet formation theory \citep{Oberg2011}, interior geochemical modelling \citep{Gaillard2014}, lunar evidence \citep{Halliday2008} and analysis of zircons \citep{Yang2014}, some clear insights into the nebular past of Earth's atmosphere emerge. As \citet{Gaillard2014} argue, it appears as though the Earth's atmosphere has gone through several phases in its 4.57 billion year history, and it is very difficult to attach accurate dates to each of these phases.

When the Earth first formed, its surface would likely have been a global magma ocean, heated from the formation process itself, constant impacts (including the moon-forming impact 4.43 billion years ago), and radioisotope decay. Its initial atmosphere would have been a nebular gas of low molecular weight, composed largely of molecular hydrogen, with trace (1000 ppm) amounts of CO, CH$_4$, NH$_3$ and N$_2$ whose relative abundances would have been determined by formation history and the surface and atmospheric temperature \citep{Lammer2011}. This phase of Earth's atmosphere may have lasted for $10^6$ to $10^8$ years according to atmospheric escape models (although there is a large amount of uncertainty from the temperature profile of the atmosphere, the sun's activity and the impact history during this time \citep{Lammer2014}.

The atmospheric loss of nebular material would have been replenished by C-H-S-rich impactors during the tail end of accretion \citep{Holzheid2000,Hashimoto2007,Zahnle2010,Brasser2016}, and would have resulted in a largely reducing atmosphere. Not much is known about this second atmosphere, and how it would have transitioned into a third, likely \ce{CO_2}-rich atmosphere. From 3.8 Gya, the global redox state of the upper mantle has remained effectively unchanged \citep{Delano2001}, but this is not a good basis for predicting the redox state of the atmosphere. The redox state of the crust would be a much better indicator for the atmospehric composition, but is much harder to constrain. There is some evidence, from \citet{Yang2014},  that the crust and atmosphere of the Earth was reducing between 4.3 and 3.6 Gya, in contrast with previous assumptions which asserted that hydrogen escaped before this time \citep{Kasting1993}, but in good agreement with recent investigations into the atmospheric escape of Xenon \citep{Zahnle2019}. The atmospheric carbon would likely have been mostly in the form of CH$_4$, CO or CO$_2$. 

In the context of such great uncertainty, exoplanets can be seen as laboratories within which to explore a diverse range of bulk and atmospheric compositions \citep{Hu2014, Rimmer2019a}. Given the diversity of exoplanet compositions and their formation histories, it is likely that some rocky exoplanets would have atmospheres where the number of carbon atoms equals or even exceeds the number of oxygen atoms (C/O > 1). For these planets, if sufficient amounts of hydrogen are also available, the bulk atmospheric composition will be primarily CO and/or CH$_4$ \citep{Hu2014,Rimmer2019b}. This provides us a considerable advantage: in principle, we can use these young exoplanetary environments as experimental environments to find out how sensitive the buildup of a prebiotic inventory is to the bulk atmospheric composition and photochemical/solar forcing.

What would impacts do to these reducing atmospheres? We answer this question by combining experiment and atmospheric modelling. For the experiment, we start in Section \ref{sec:experiment} with a reducing 1 bar gas containing equimolar quantities of methane, carbon monoxide and nitrogen along with water in both liquid and vapor form. This gas is exposed to a laser pulse, simulating a high-energy impact, and the resulting chemical yields are measured spectroscopically. The yields are then incorporated in Section \ref{sec:photochem} as outgassing fluxes from the planet's surface within a photochemical model that predicts the atmospheric chemistry during a phase of heavy bombardment, either a `late heavy bombardment', or bombardment by impacts at the the tail of accretion, a stage that can substantially change the atmosphere \citep[see, e.g.][]{Wyatt2019}. The resulting transmission spectrum for exoplanets with this chemistry are modeled as described in Section \ref{sec:trans-model}. The results for the experiments, atmospheric model and transmission spectra are presented in Section \ref{sec:results}, and are then discussed (Section \ref{sec:discussion}).

\section{The PALS Laser Experiment}
\label{sec:experiment}
\begin{figure}
\centering
\includegraphics[width=\textwidth]{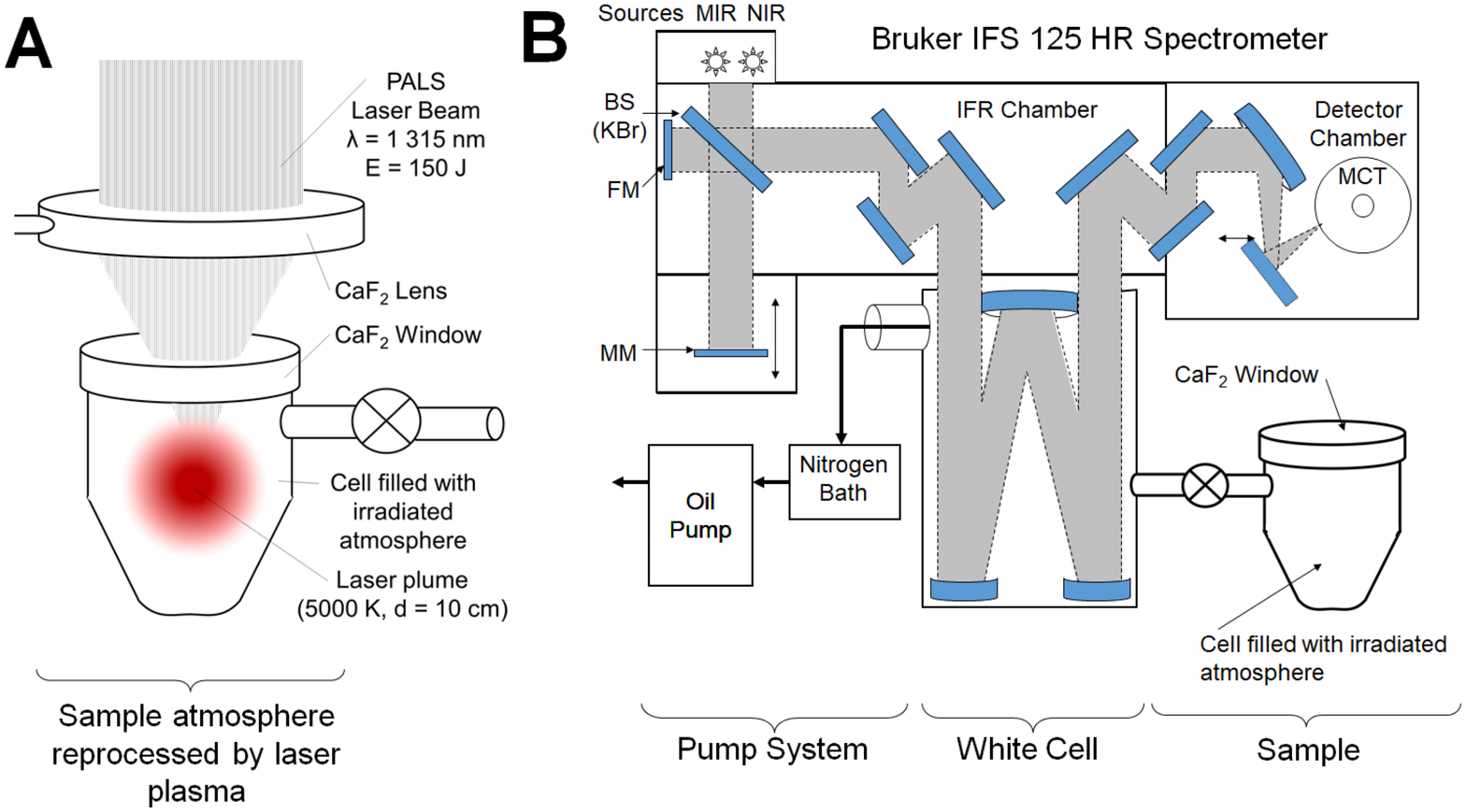}
\caption{Panel A shows a schematic diagram of the experimental arrangement shows a set-up used in the Terawatt Laser Facility. MM indicates "moving mirror" of the interferometer (IFR), FM "fixed mirror," BS the beamsplitter (Potassium Bromide). A laser beam is focused by a lens to the center of an interaction cell filled with a mixture representing the particular atmosphere studied in our experiment. Panel B shows optical set-up of our FTIR (Fourier Transform Infrared) spectrometer equipped with a multipass cell. The interaction cell is transferred from the laser facility, connected to the multipass cell at which point a sample of gas is taken. Then, an infrared spectral survey of the gas composition is performed.\label{fig:scheme}}
\end{figure}

The reprocessing of an CH$_4$ + CO + H$_2$O atmosphere was initiated by a laboratory simulated impact shock. The  simulation of the high-velocity impact of an extra-terrestrial body into the early planetary atmosphere was performed due to a high-power  laser induced  dielectric breakdown (LIDB) of the gaseous mixture at the  Prague Asterix Laser System (PALS) facility. Simple molecular products were monitored by means of infrared absorption spectroscopy. Fig. \ref{fig:scheme} gives a schematic diagram of the experiment.

An equimolar gaseous mixture of CH$_4$ + CO + N$_2$ representing an atmosphere at a nominal pressure of 760~Torr was exposed to 25 laser pulses focused to induce a hot, dense plasma, simulating high-velocity impact conditions. Liquid water (1 mL) and montmorillonite powder were added to the system. In the gas phase, water vapor pressure was saturated (23.8 Torr). There were several reference samples irradiated under the same conditions but without any presence of the solid material. \citet{Brederlow1983} and \citet{Jungwirth2001} show that laser pulses were delivered by a high-power (0.4~TW) iodine photo-dissociation laser system PALS (Prague Asterix Laser System with a pulse duration of 350~ps, and a wavelength of 1.315~$\mu$m). A single pulse carried the energy of 150 J. One pulse was generated every 20 minutes, and more than 8 hours were required to conduct the experiment. The laser beam was focused in the centre of the thick-walled glass cell by a plano-convex lens with a diameter of 15 cm and a focal length 25~cm. Pulse energy losses at the focusing lens and cell window did not exceed 15 \%. A hot, dense plasma was formed in the cell by laser-induced dielectric breakdown (LIDB). The centimetre-sized plasma fireball can be considered as a good laboratory model of a high-velocity impact and/or lightning in planetary atmospheres \citep{Babankova2006,Juha2008}. This is because, during its spatiotemporal evolution, the fireball passes through various stages relevant to these high-energy-density natural phenomena. Shock and thermal waves as well as energetic photons and charged particles are emitted from the hot core of the laser spark. The expanding plasma then undergoes fast quenching by mixing with the ambient gas; a frozen equilibrium takes place in the chemistry of the system after every laser shot. 

Well after the reactions have completed, the composition of the gas phase in the cell was monitored after the aforementioned laser pulses, using the Bruker IFS 125 HR spectrometer (Bruker Optics, Germany) equipped with a KBr beam splitter and a nitrogen cooled MCT detector over the spectral range of 600 to 6000 cm$^{-1}$. The spectra were measured with a resolution of 0.02 cm$^{-1}$ using the Blackmann-Harris apodization function \citep{Harris1978}. The sample was transferred to a White multi-pass cell with an optical beam length of 10~m using a vacuum line. 300 scans were accumulated for each measurement.  Concentrations of all the gases were determined by independent calibration measurements of pure standards. Integrated intensities of randomly selected individual absorption lines  of each species were calculated using the OPUS 6.0 software package, and the data were subsequently manually fitted by a linear regression model. 

\section{The Photochemistry/Diffusion Model}
\label{sec:photochem}
We use the volatile concentrations obtained from the laser experiment as input for the atmosphere both for the climate and photochemical models. The abundances: 10\% CO, 10\% \ce{CH_4}, 80\% \ce{N_2}, surface \ce{H_2O} at vapor pressure are taken to be the bulk atmospheric composition for an atmosphere with a fixed surface pressure of 1 bar. We apply the above composition to calculate the temperature profile using the ATMOS 1-D Global Climate model (Section \ref{sec:initial-conditions}). The impact results obtained as described in Section \ref{sec:experiment} are then used to estimate the surface atmospheric mixing ratios (Section \ref{sec:outgassing-rates}). The photochemistry is solved in Section \ref{sec:photochem-model} using the STAND2018 network and the ARGO photochemistry-diffusion model. The bulk atmospheric composition that we use for the climate and photochemistry models is different from the gas composition used for the experiment. The effect that atmospheres of different compositions on impact-generated chemistry is discussed in Section \ref{sec:bulk-chemistry}.

\subsection{Planetary Climate and Temperature Profile Used in this Research}
\label{sec:initial-conditions}

The initial and boundary conditions discussed above are applied to calculate the temperature profile using the ATMOS 1-D Global Climate model \citep{Kasting1984,Pavlov2000,Haqq2008,Koppa2013,Ramirez2014}.

\begin{figure}
\centering
\includegraphics[width=0.48\textwidth]{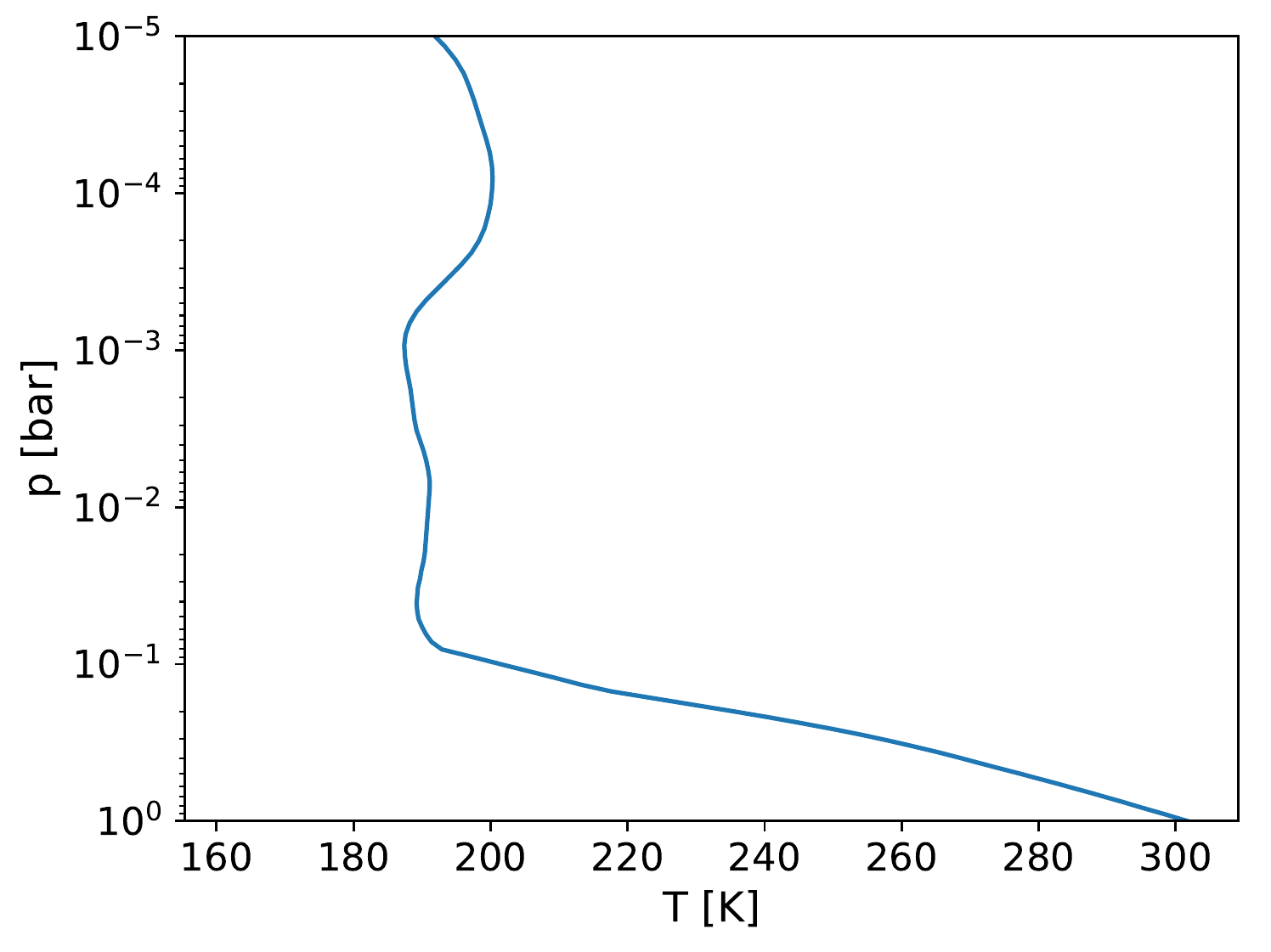}
\includegraphics[width=0.48\textwidth]{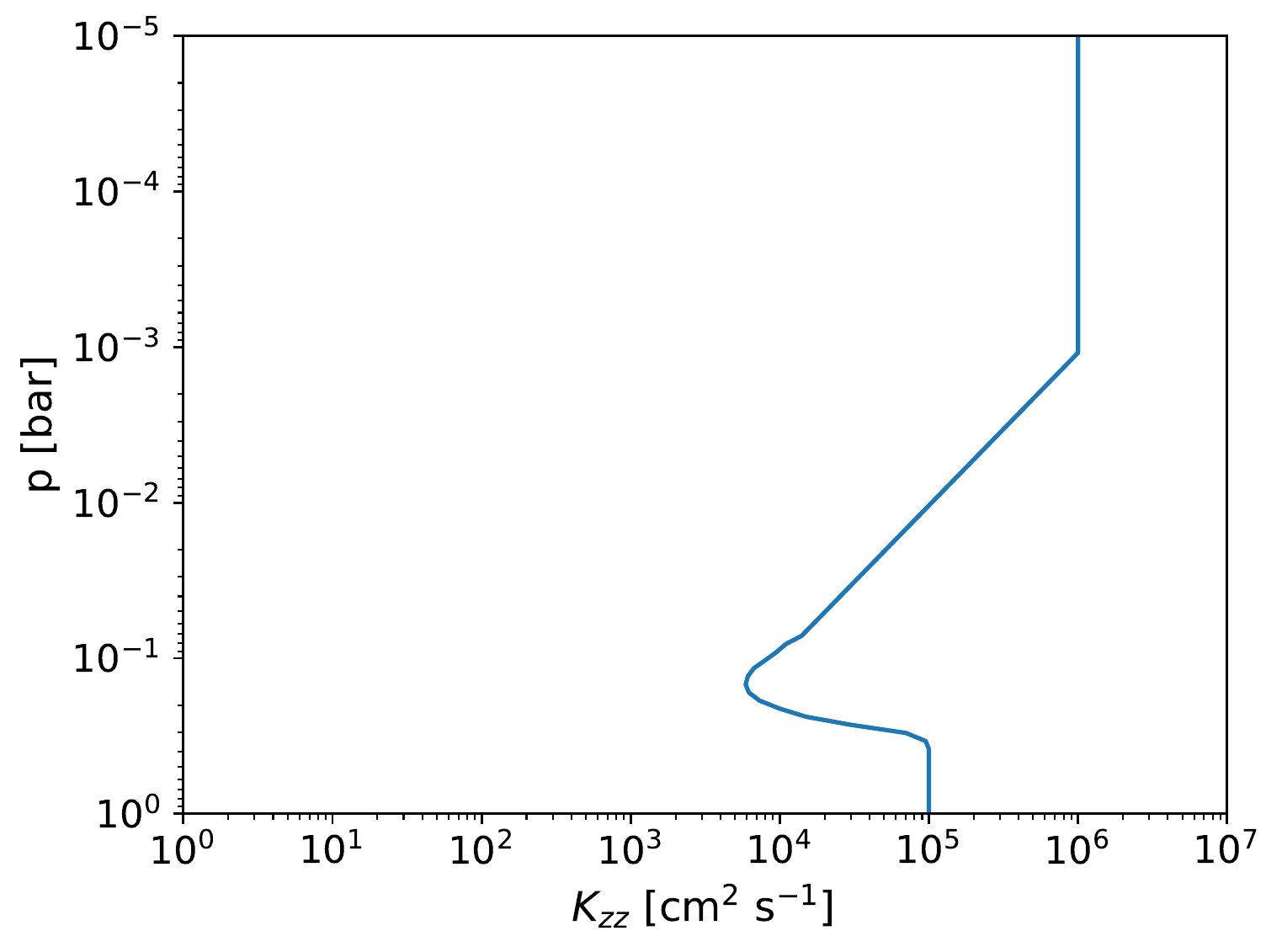}
\caption{Temperature [K] (left) and eddy diffusion ($K_{zz}$ [cm$^2$ s$^{-1}$], right), both as a function of atmospheric pressure ($p$ [bar]). }
\label{fig:t-profile}
\end{figure}

The relevant greenhouse gases are \ce{H_2O} and \ce{CH_4}. \ce{CH_4} is a very potent greenhouse gas, and at such high abundances, by itself, might not result in a temperate surface environment. At the same time, abundant \ce{CH_4} is expected to produce a photochemical haze \citep{Trainer2006}, which will cool the planet's surface. This interplay, along with the surface heat flux due to impacts and latent heat from planet formation make it very difficult to constrain the temperature. Because of this complex interplay, we treat \ce{CH_4} as \ce{CO_2} for the purposes of the climate model, to mitigate the greenhouse effect, and leave for the future a more accurate climate simulation that accounts for the many factors present on an Earth-Like planet experiencing heavy bombardment. It is worth noting that the chemistry near the surface is not especially sensitive to the temperature profile in the sense that acetylene, hydrogen cyanide and ammonia destruction even at 100 $^{\circ}$C is nevertheless dominated by photochemistry and diffusion. The temperature and eddy diffusion profiles are shown in Figure \ref{fig:t-profile}.

\subsection{Estimating the Outgassing Rates and Surface Mixing Ratios of Impact-Generated Species}
\label{sec:outgassing-rates}

The abundances of the shock chemical products correspond to exposure to a laser blast of a certain energy, and gives us a number of molecules per Joule of energy ($S_i$, for species $i$). We incorporate the experimental results in terms of surface fluxes, and preserve mass balance of the non-condensables by sequestering excess hydrogen and oxygen into water.

The surface flux can be expressed in terms of the average of this source function, the energy deposited via an impact, $E_i$, and the frequency of impacts $\nu_i$: $\langle E_i \nu_i \rangle$. This product is considered to be evenly distributed over the surface of the planet, so the product is divided by the surface area of the planet to give the surface flux from the impact, $\Phi_{s,i}$ [cm$^{-2}$ s$^{-1}$]:
\begin{equation}
\Phi_{s,i} = \dfrac{S_i \langle E_i \nu_i \rangle}{4\pi R_p^2}.
\label{eqn:flux}
\end{equation}
The average of the energy deposition and frequency can be expressed:
\begin{equation}
\langle E_i \nu_i \rangle = \int_{E_0}^{\infty}\nu(E) \; dE.
\label{eqn:moment}
\end{equation}
The impact frequency is expressed in terms of crater diameter, $D$ [m], as a power law: \citep{Wetherill1975}:
\begin{equation}
\nu = \nu_0 \, \Bigg(\dfrac{D}{D_0}\Bigg)^{\!\!-\alpha}
\label{eqn:freq}
\end{equation}
Where $D_0 = 10^4$ m is a reference diameter, $\nu_0 = \nu(D_0)$ and we take $\alpha = 1.3$, which seems to agree with the lunar crater size distribution reasonably well. The diameter of the crater, in turn, can be related to the energy, also as a power-law \citep{Wetherill1975}, with constants taken from the literature \citep{Hughes2003}:
\begin{equation}
D = 4.8 \times 10^{-6} \, \text{m} \; \Bigg(\dfrac{E}{\text{1 erg}}\Bigg)^{\!\!1/\beta}, 
\label{eqn:diameter}
\end{equation}
where $\beta$ can be between 3 (energy limited) or 4 (gravity limited) \citep{Wetherill1975}. We choose $\beta = 3$. From the relationship between crater diameter and impactor mass from \citep{Wetherill1975}, we can parameterize the impactor frequency as a function of the mass deposition of impactors:
\begin{equation}
\nu_0 = \dfrac{\dot{M}_T}{M_0},
\label{eqn:freq0}
\end{equation}
Where $\dot{M}_T$ [g/s] is the mass deposition rate and $M_0 = 6.6 \times 10^{17}$ g. Applying Eq's (\ref{eqn:freq}) - (\ref{eqn:freq0}) to Eq. (\ref{eqn:moment}), applying a high-energy cutoff for the impactor, $E_{\rm max} = 10^{36}$ erg (assuming the maximum crater size of 1000 km and applying Eq. (\ref{eqn:diameter})). We then integrate to find:
\begin{equation}
\langle E_i \nu_i \rangle = \dfrac{\dot{M}_T}{M_0} \dfrac{E_0}{1-\gamma}\Bigg(\dfrac{E_{\rm max}}{\text{1 erg}}\Bigg)^{\!\!1-\gamma},
\label{eqn:moment-solved}
\end{equation}
where $E_0 = 1.3 \times 10^{12}$ erg and $\gamma = \alpha/\beta = 0.43$.  
We apply this result Eq. (\ref{eqn:flux}), along with $S_i$ [molecules/J] from Table \ref{table:source}, taking our mass deposition rate during the LHB to be $\dot{M}_T = 4.3\times 10^{13}$ g/year. 

\begin{deluxetable}{c|ccc} 
\centering
\tabletypesize{\footnotesize}
\tablewidth{290pt}
\tablecolumns{12} 
\tablewidth{0pt} 
\tablecaption{Experimental yield of species, corresponding surface fluxes and atmospheric lifetimes. \label{table:source}} 
\tablehead{
\colhead{Molecule} & \colhead{$S_i$ [mol/J]} & \colhead{$\Phi_{i,s}$ [cm$^{-2}$ s$^{-1}$]} & \colhead{$\tau$ [yr]}} 
\startdata
\ce{HCN} & $5.4 \times 10^{-7}$ & $9.7 \times 10^{12}$ & 30\\ 
\ce{C_2H_2} & $8.6 \times 10^{-7}$ & $1.6 \times 10^{13}$ & 250\\ 
\ce{NH_3} & $1.1 \times 10^{-7}$ & $1.9 \times 10^{12}$ & 15 \\ 
\enddata
\tablecomments{Values for $S_i$ are based on experimental results detailed here, values for $\Phi_{i,s}$ are from solving Eq. (\ref{eqn:flux}) with Eq. (\ref{eqn:moment-solved}), and lifetimes derived from our exoplanet model atmosphere using the STAND network and ARGO chemical model for photodestruction rates (See Section \ref{sec:after-impact}).}
\end{deluxetable}

\subsection{The ARGO Photochemistry/Diffusion Model}
\label{sec:photochem-model}

The surface fluxes from Table \ref{table:source} are incorporated into ARGO, a Lagrangian photochemistry and cosmic-ray atmospheric chemistry model \citep{Rimmer2016}, that takes a prescribed temperature profile, which we determine using a climate model (discussed above), a high resolution (1~\AA) UV field estimated  for the 1~Gy sun \citep{Ribas2005,Ribas2010}, and a comprehensive chemical network, STAND \citep{Rimmer2019b}, valid between 300~K and 30000~K incorporating H/C/N/O ion and neutral chemistry including complex hydrocarbons and amines, including the amino acid glycine \citep{Rimmer2016}. The model solves the equation:
\begin{equation}
\dfrac{d[\ce{X}]}{dt} = P(\ce{X}) - L(\ce{X})[\ce{X}] - \dfrac{\partial \Phi(\ce{X})}{\partial z},
\end{equation}
where [\ce{X}] (cm$^{-3}$) is the concentration of chemical species \ce{X}, itself a function of the atmospheric height, $z$. The production of \ce{X} is denoted by $P(\ce{X})$ [cm$^{-3}$ s$^{-1}$] and loss by $L(\ce{X})$ [s$^{-1}$], and the flux due to diffusion by $\Phi(\ce{X})$ [cm$^{-2}$ s$^{-1}$].

The model is Lagrangian, and keeps track of the time-scales at which the parcel exists at a particular temperature  and pressure, and then changes the temperature and pressure. Molecular diffusion is approximated using `banking' reactions. This model can be applied to planetary atmospheres from hot Jupiters and hot Super Earths, to Earth, and is even reasonably accurate for Jupiter \citep{Rimmer2016}. We use the model with corrected downward diffusion \citep{Rimmer2019err}.

In order to account for gradual geological emissions of gases at the surface of the planet, a new outgassing process was added to the model. A new type of reaction was specified within the STAND network, which only takes place at the surface. This reaction converts a new `ground' species, \ce{GX}, available at virtually infinite abundance, but not contributing to any other process, to the chemically active species. The rate of outgassing is controlled by the rate of this conversion as well as the time that the parcel spends at the surface. The rate is set to:
\begin{equation}
R_{\rm out}(\ce{X}) = \dfrac{\Phi(\ce{X})}{\Delta z},
\end{equation}
where $\Phi(\ce{X})$ [cm$^{-2}$ s$^{-1}$] is the surface emission flux, $\Delta z$ [cm] is the height step of the atmospheric model, and $R_{\rm out}(\ce{X})$ [cm$^{-3}$ s$^{-1}$] is the effective rate at which species \ce{X} is introduced into the atmosphere, and this last quantity is incorporated into the model.

For the network, we have updated the rate constants for some reactions, namely:
\begin{align}
\ce{O} + \ce{O} + \ce{M} &\rightarrow \ce{O_2} + \ce{M}, & k_0 = 1.67 \times 10^{-33} \, \Big(T/300 \, {\rm K}\Big)^{-1}, \label{R1}\\
\ce{O} + \ce{OH} &\rightarrow \ce{O_2} + \ce{H}, & k = 3.28 \times 10^{-11} \, \Big(T/300 \, {\rm K}\Big)^{-0.32} , \label{R2} \\
\ce{C_2H_2} + \ce{H} + \ce{M} &\rightarrow \ce{C_2H_3} + \ce{M}, & k_{\infty} = 3.01 \times 10^{-11}\Big(T/300 \, {\rm K}\Big)^{1.09} \, e^{-1327/T},\label{R3} \\
\ce{C_2H_2} + \ce{CH_3} + \ce{M} &\rightarrow \ce{C_3H_5} + \ce{M}, & k_{\infty} = 6.09 \times 10^{-9}\, \Big(T/300 \, {\rm K}\Big)^{-5.98} \, e^{-6695/T}, \label{R4}\\
\ce{HCN} + \ce{C_2H_3} &\rightarrow \ce{C_3H_3N} + \ce{H}, & k_{\infty} = 1 \times 10^{-12} \, e^{-900/T}. \label{R5}
\end{align}
Where Reaction (\ref{R1}) is from \citet{Javoy2003}, Reaction (\ref{R2}) is from \citet{Robertson2006}, Reaction (\ref{R3}) is from \citet{Knyazev1996}, Reaction (\ref{R4}) is from \citet{Diau1994}, and Reaction (\ref{R5}) is from \citet{Monks1993}. The value NIST provides for Reaction (\ref{R5}) is incorrect; they quote the 298 K value as though it is constant over all temperatures, whereas \citet{Monks1993} gives the rate constant as above. We have also removed the reaction:
\begin{equation}
\ce{C_2H_5} + \ce{HCO} + \ce{M} \rightarrow \ce{C_3H_6O} + \ce{M},
\end{equation}
because the published rate constant was extracted indirectly from an experiment where atomic hydrogen was reacted with \ce{C_2H_4} and \ce{CO}, and it is unclear whether the measured \ce{C_3H_6O} resulted from the above reaction. In addition, we have updated the absorption cross-sections for several photochemical reactions using the MPI-Mainz UV/VIS spectral atlas \citep{Keller2013}.

\section{Models of the Transmission Spectra}
\label{sec:trans-model}
The line lists of H$_2$O \citep{06BaTeHa.H2O}, CO \citep{15LiGoRo}, CO$_2$ \citep{HITEMP2010}, CH$_4$ \citep{14YuTexx.CH4}, HCN \citep{14BaStHi.HCN}, NH$_3$ \citep{11YuBaTea} and C$_2$H$_2$ \citep{17LyPe.C2H2} were used to build the temperature and pressure dependent cross sections as described in  \citet{ExoCross}.  The line lists are taken from the ExoMol \citep{ExoMol,16TeYuAl}, HITRAN \citep{HITRAN2016}, HITEMP \citep{HITEMP2010} and ASD-1000 databases \citep{17LyPe.C2H2}. 

We used an adaptive wavenumber grid ranging from 2$\times10^{-4}$ cm$^{-1}$ at the longest wavelength (15\,$\mu$m) and 0.01 at the shortest wavelength (0.5\,$\mu$m) considered. The grid of temperatures ranges from 200~K to 2000~K (by 100~K), while the grid of pressures is  distributed in the $\log$-space from from 0.0001~atm to 50~atm. We calculate the synthetic transmission spectra using the forward model of the open-source retrieval framework TauREx \citep{Waldmann2015}. We assume a plane-parallel atmosphere with 100 vertical layers and include conditional induced absorption due to hydrogen and helium as well as Rayleigh scattering contributions for H$_2$O, H$_2$,He, N$_2$, O$_2$, CO$_2$, CH$_4$ and NH$_3$. We model a cloud-free, clear atmosphere.

\section{Results}
\label{sec:results}

An equimolar gas mixture of CH$_4$ + CO + N$_2$ is reprocessed by laser pulses, representing the shock save and plasma created in planetary atmosphere by an asteroid impact. These results were examined by high resolution FTIR spectrometry (Section \ref{sec:plasma-chem}) and the results are combined with modeled atmospheric chemical profiles (Section \ref{sec:atmospheric-chemistry}) From the atmospheric chemistry, the evolution of surface chemistry as impactor frequencies decline is shown (Section \ref{sec:impact-composition}), and exoplanetary transition spectra are simulated (Section \ref{sec:transmission}). Our explorations show that namely acetylene detected spectroscopicaly among the main products can be detected.  

\subsection{Laser-Induced Plasma Chemistry}
\label{sec:plasma-chem}

Experimental results show that amounts of C$_2$H$_2$, HCN, and NH$_3$, significant for atmospheric chemistry, are generated in the impact simulation, transforming the atmosphere (see Table \ref{table:results1}). Our spectroscopic analysis shows that the experimental atmosphere containing equimolar ratios of CO, CH$_4$, N$_2$ and water is reprocessed, yielding several products, the dominant species among them being acetylene (C$_2$H$_2$) and hydrogen cyanide (HCN) in mutual ratio of approximately 3:2, accompanied with a less than a few percent ammonia, ethylene, and carbon dioxide. Molecular hydrogen is generated mainly during decomposition of methane and we estimate that each mole of methane ends up producing about 0.6~mol of H$_2$ (the remainder is retained in the products). Because the molecular hydrogen cannot be directly measured, and because it is expected to rapidly escape, we do not include it in the atmospheric model.

\startlongtable
\begin{deluxetable}{ccc|cccc|ccc} 
\centering
\tabletypesize{\footnotesize}
\tablewidth{290pt}
\tablecolumns{12} 
\tablewidth{0pt} 
\tablecaption{Experiment and Model Compositions and Properties, Compared to Atmospheres of Rocky Bodies  \label{table:results1}} 
\tablehead{ 
 \colhead{Environment} & \colhead{$p$} & \colhead{$T$} & \multicolumn{4}{c}{Bulk Composition} & \multicolumn{3}{c}{Trace/Other Gases} \vspace{-0.2cm}\\
 & \colhead{[bar]} & \colhead{[K]} & \colhead{N$_2$} & \colhead{CO} & \colhead{CH$_4$} & \colhead{CO$_2$} & \colhead{HCN} & \colhead{C$_2$H$_2$} & \colhead{NH$_3$} } 
\startdata
Exp Start\tablenotemark{*} & 1.0 & 295. & 33\% & 33\%   & 33\%  & 0\%   & 0\%       & 0\%       & 0\%    \\ \hline
Exp Finish\tablenotemark{$\dagger$} & 1.0 & 295. & 37\% & 30\%   & 11\%  & 1\%   & 5\%       & 8\%   & 1\%   \\ \hline
Model Atmosphere\tablenotemark{$\ddagger$}  & 1.0 & 295. & 80\% & 10\%   & 10\%  & 0\%   & 400 ppm   & 0.4\%  & 40 ppm   \\ \hline
Early Earth\tablenotemark{a} & 1.0 & 295. & 99.6\% & 200 ppm & 0.1\% & 0.3\%  & 50 ppb     & 0\% & $< 50$ ppb \\ \hline
Modern Volcano\tablenotemark{b} & &  &  0 -- 1\% & 0 -- 2.4\% & 0 -- 99\% &  0 -- 99\%   &           &  0 -- 5\%  &  $< 1$ ppb \\ \hline
Hadean Volcano\tablenotemark{c} & 10 & 1500 &  0 -- 0.2\% & 1 -- 25\% & 0 -- 0.2\% & 0 -- 50\%   &   0 -- 500 ppm     &  0 -- 10\%  & 0 -- 1 \%  \\ \hline
    Titan\tablenotemark{d} & 1.5 & 94.  & 94\% & 10 ppm & 5.7\% & 1 ppb & 2 ppm     & 2 ppm & $< 1$ ppb
\enddata
\tablenotetext{*}{Initial experimental mixture. A small amount of liquid water was also included, to provide water vapor after the experiment was performed.}
\tablenotetext{\dagger}{Experimental mixture after 3750 J deposit of energy. See Section \ref{sec:experiment} for details.}
\tablenotetext{\ddagger}{The outcome of applying experimental results as surface fluxes and solving for the photochemistry.}
\tablenotetext{a}{Based on models by \citet{Tian2011}}
\tablenotetext{b}{Observations of volcanic plume chemistry span these ranges, and also span a wide range of degassing temperatures and pressures; see \citet{Fischer2008,Hedberg1974,Etiope2004}.}
\tablenotetext{c}{Ranges come from models of Early Earth volcanic plumes over a wide range of oxygen fugacities \citep{Rimmer2019a}.}
\tablenotetext{d}{Concentrations either observed or inferred from experiment, see \citet{Niemann2010,Teanby2007,Horst2008}}
\tablecomments{Values given are experiments or observations where possible. The Early Earth values are global averages from a model informed by various lines of geological evidence \citep{Tian2011}. The values for modern and early volcanoes are ranges. Modern volcano values are based on measurements. Early volcano values are taken from a model \citep{Rimmer2019a}.}
\end{deluxetable}

HCN in particular is well-connected to prebiotic chemistry, and is invoked in a host of different scenarios \citep{Ferris1978,Ritson2012,Xu2018}. We apply the experimental chemical yields of C$_2$H$_2$, HCN, and NH$_3$ to a photochemistry/diffusion model in order to predict the atmospheric signatures of impacts. Specifically, we apply these results to a model atmosphere for an Earth-like exoplanet (Earth mass, Earth radius, 1 AU away from a Sun-like star), with a bulk chemistry dominated by 80\% N$_2$, 10\% CO, and 10\% CH$_4$ in the presence of surface liquid water and its saturated vapour. We include the impact-generated \ce{C_2H_2}, \ce{HCN} and \ce{NH_3} as surface fluxes, as described in Section \ref{sec:outgassing-rates}. The temperature profile and other details of this atmosphere are discussed in Section \ref{sec:initial-conditions}.

\subsection{Atmospheric Chemistry}
\label{sec:atmospheric-chemistry}

Here we discuss the chemistry of an Earth-like rocky exoplanet with the bulk atmosphere of 80\% \ce{N_2}, 10\% \ce{CO} and 10\% \ce{CH_4}, with water at vapor pressure. We assume this atmosphere is stable, and its stability is largely justified by indications from the geological record that Earth may have had this atmospheric composition during the Hadean \citep{Yang2014}. If the planet is lifeless, then high concentrations of \ce{CO} can easily be maintained. High surface fluxes of \ce{CH_4} could lead to its comprising 10\% of the atmosphere, especially if the escape of \ce{H_2} is slow. We first show the photochemistry without impacts.

The main consequence of the photochemistry before impacts is in the fixing of nitrogen high in the atmosphere, and the generation of large hydrocarbons, that probably develop into an upper-atmospheric haze. This is consistent with experimental and other model results for gases with high concentrations of \ce{CH_4} \citep{Trainer2004,Trainer2006,Arney2016}, and that CO changes the character of the resulting haze \citep{Horst2014}. Significant amounts of hydrocarbons are also generated in the upper atmosphere, with \ce{C_2H_2} ranging from 1 ppm at 1 mbar, up to 1000 ppm at $< 10$ $\mu$bar. For this atmosphere, \ce{C_2H_6} is more than an order of magnitude more abundant throughout. This particular atmosphere has a large fraction of highly-reduced carbon (-4, \ce{CH_4}), whereas the nitrogen is neutral (+0, \ce{N_2}), largely owing to the instability of \ce{NH_3} in this environment. This combination leads to novel photochemistry for the fixed nitrogen, with the dominant species being diazomethane (\ce{CH_2N_2}), rather than hydrogen cyanide (\ce{HCN}). The pathway for photochemical diazomethane formation is given in the Discussion (esp. Section \ref{sec:after-impact}). It would be very difficult to observe any of these fixed nitrogen species in transmission on an exoplanet with present instruments, although this may be possible with future observational capabilities \citep{Rugheimer2019}.

\begin{figure}
\includegraphics[width=\linewidth]{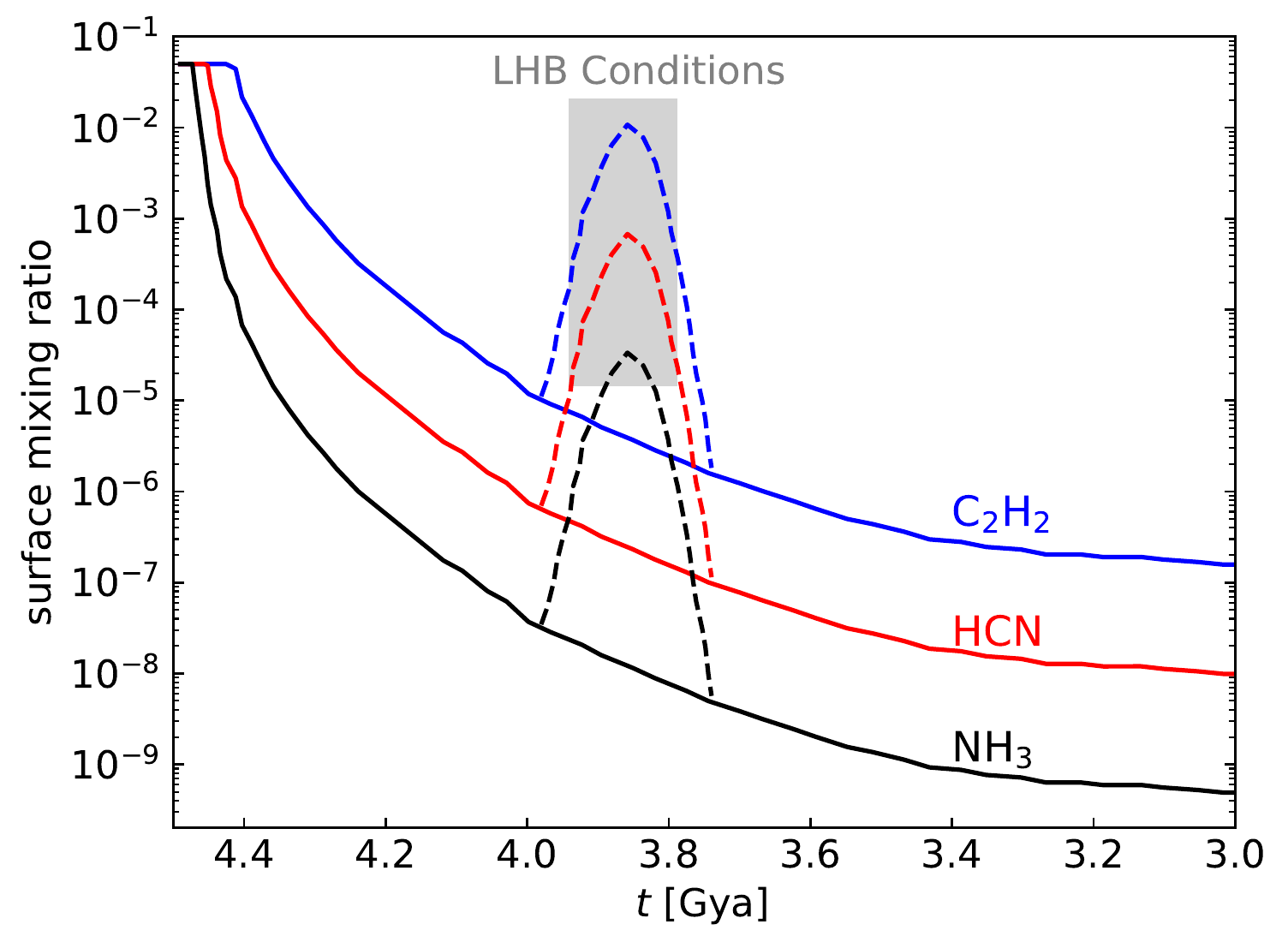}
\caption{Predicted steady-state surface mixing ratios of \ce{C_2H_2}, HCN and \ce{NH_3} over Earth's early history, formed by the impactors of an estimated frequency and energy distribution, balanced by the subsequent photochemical destruction. \label{fig:impact-time}}
\end{figure}

\begin{figure}
\centering
\includegraphics[width=0.48\textwidth]{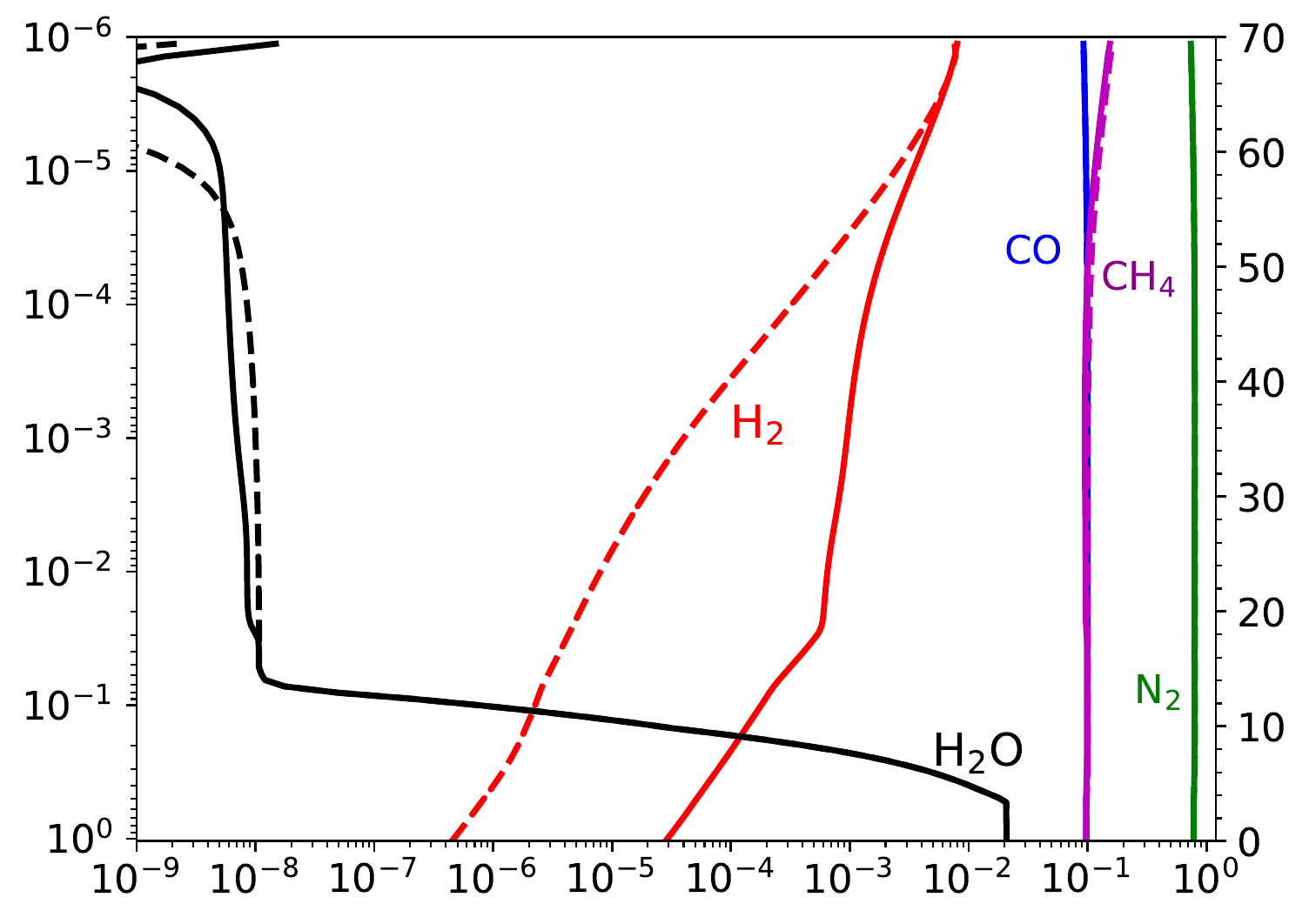}
\includegraphics[width=0.48\textwidth]{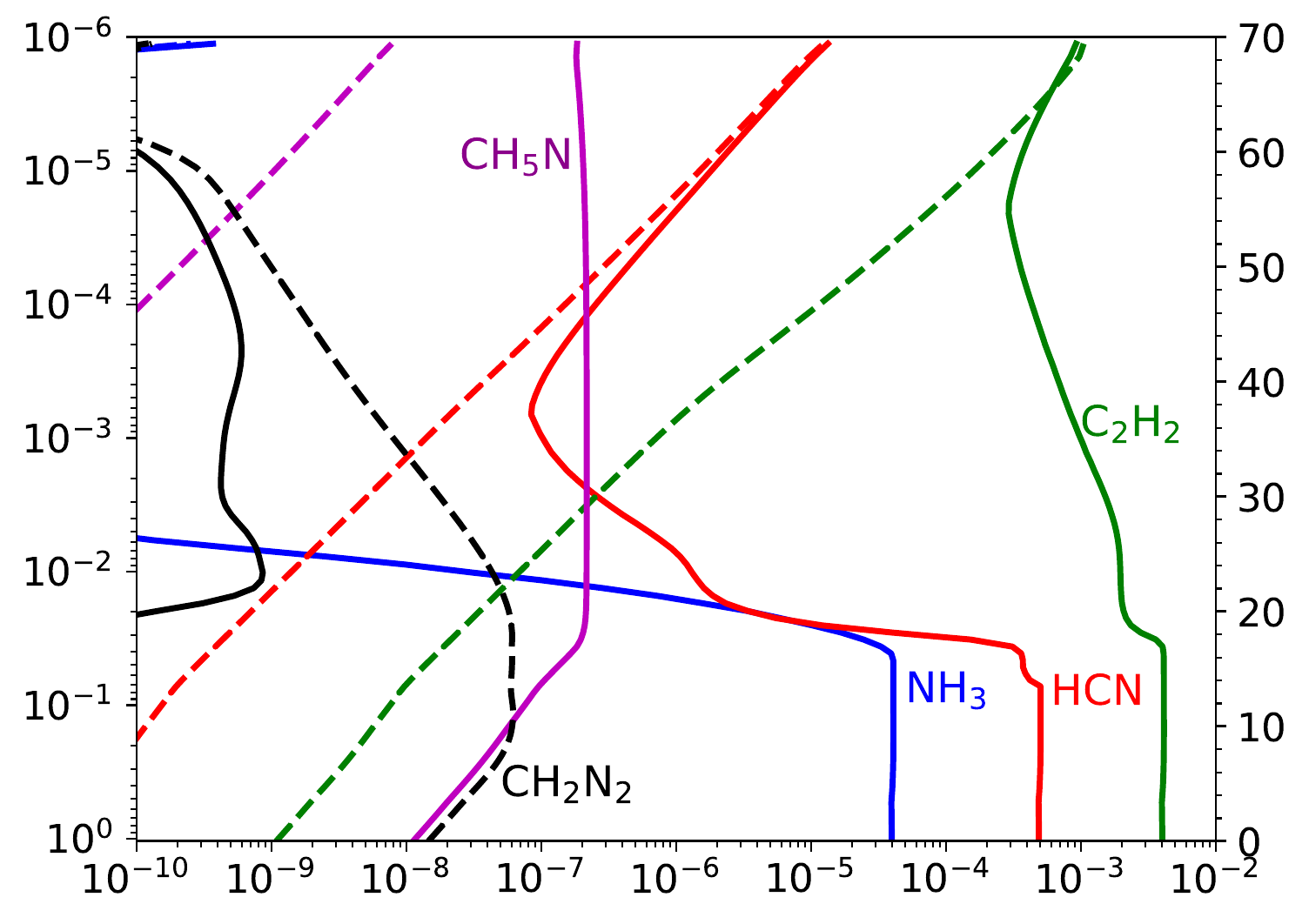}\\
\includegraphics[width=0.48\textwidth]{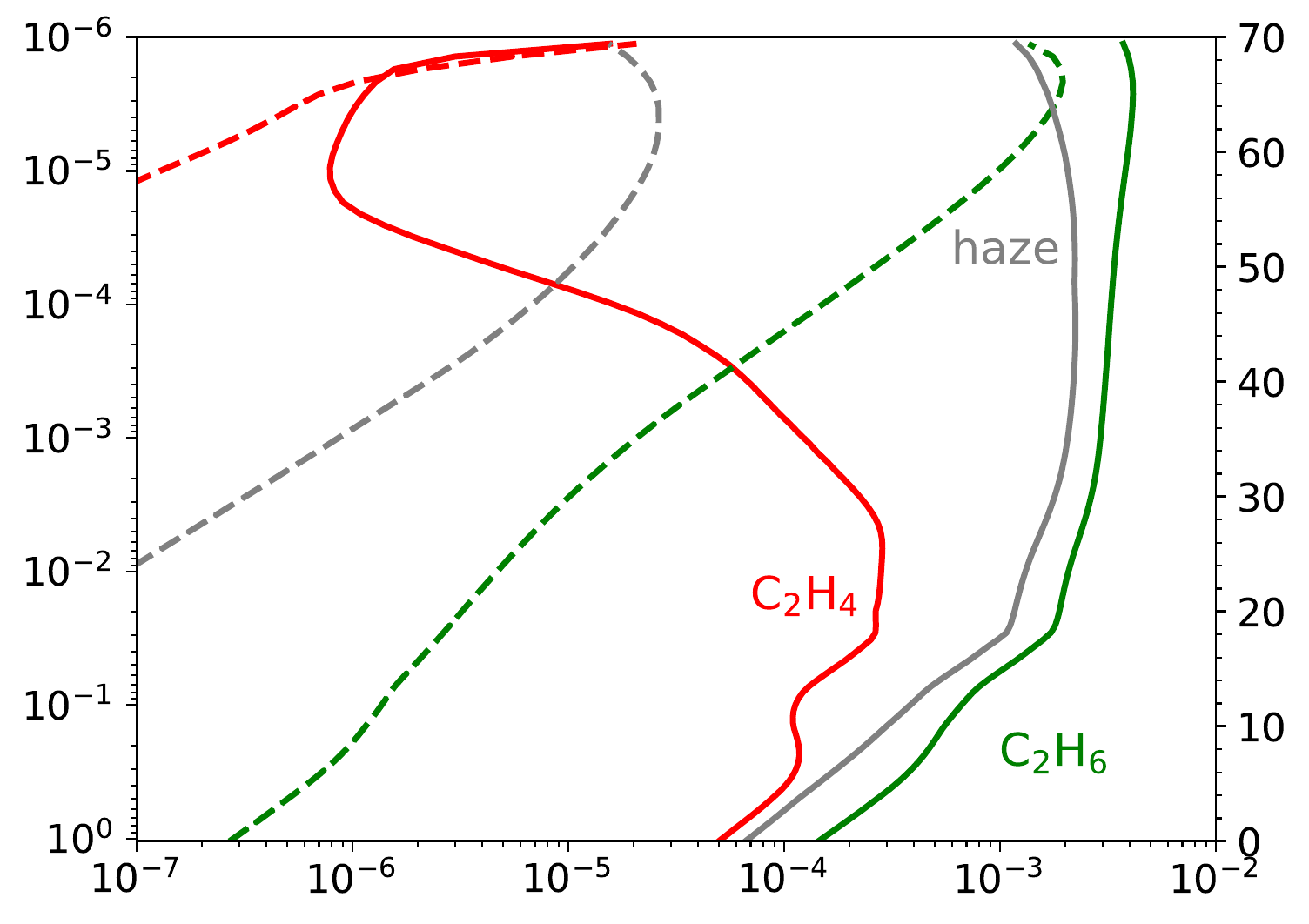}
\includegraphics[width=0.48\textwidth]{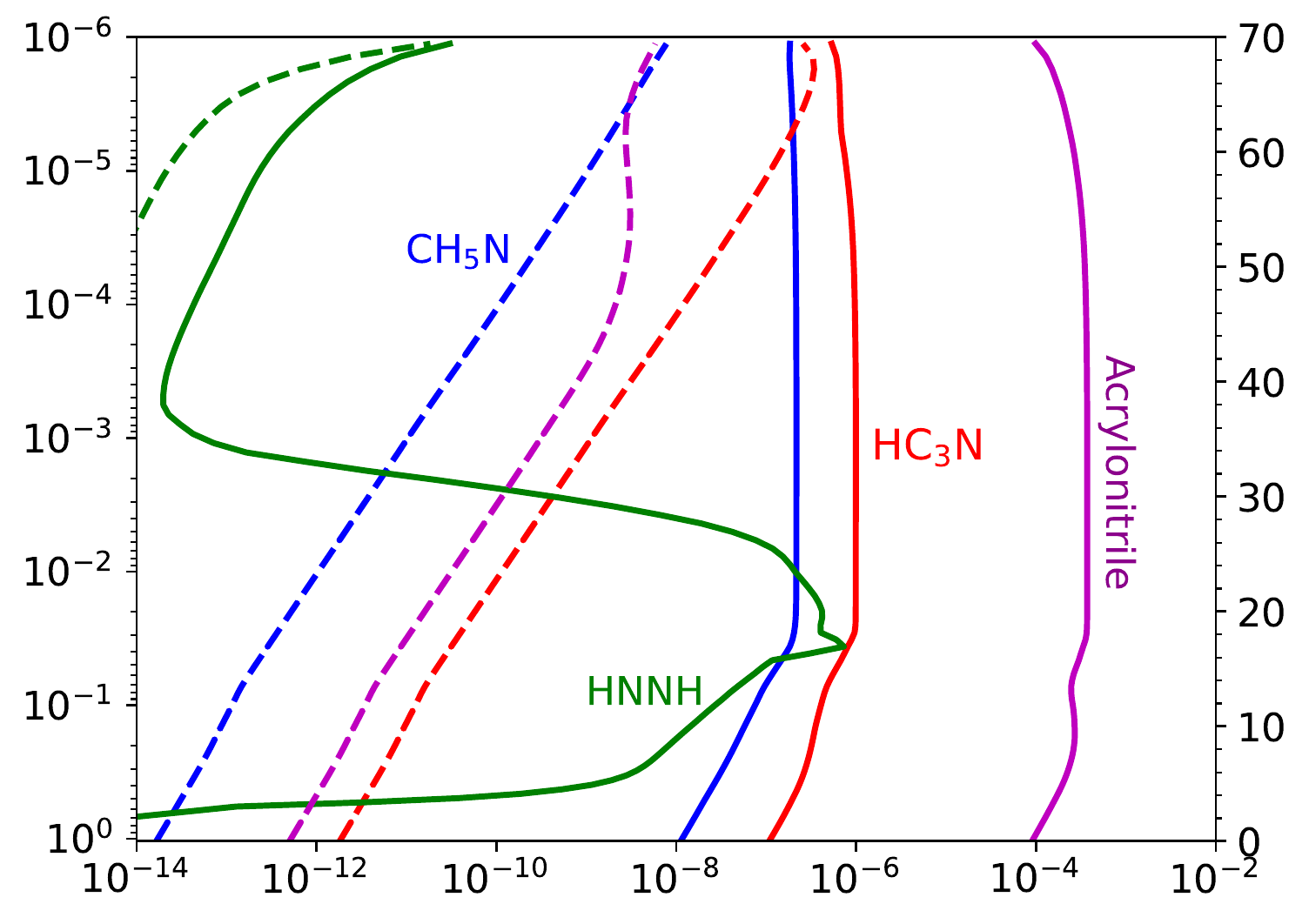}
\caption{Mixing ratios of species produced in large quantities via impact and subsequent photochemistry within a reducing temperate atmosphere (\ce{N_2}/\ce{CH_4}/\ce{CO}), and their photochemical products, as a function of atmospheric pressure $p$ [bar] (solid lines). Results for a reducing atmosphere with photochemistry but no impacts are also shown (dashed lines).}
\label{fig:photochem}
\end{figure}

Following the calculations described above (Section \ref{sec:outgassing-rates}), we incorporate the surface fluxes of C$_2$H$_2$, HCN, and NH$_3$. The dependence of the surface mixing ratios resulting from these fluxes, for different impact rates, is shown in Fig. \ref{fig:impact-time}. The resulting atmosphere is shown in Fig. \ref{fig:photochem}. We also plot the atmosphere without the impact-generated species for comparison. It is clear that photochemistry alone cannot explain the high quantities of tropospheric \ce{C_2H_2}, \ce{HCN} and \ce{NH_3}. The main effect of the photochemistry and diffusion is found in the upper atmosphere, where C$_2$H$_2$ is converted to C$_2$H$_6$ and HCN is converted to acrylonitrile. A trace amount of carbon dioxide is also produced at an atmospheric height of 30 mbar (altitude about 24 km). In addition, we predict that a high rate of impacts will generate a thick haze in the upper atmosphere. \citet{Kawashima2018} has shown that this hydrocarbon haze may obscure some of the molecular features, and will affect the surface UV actinic flux, the effects of which can be predicted \citep{Wolf2010,Arney2016}.

\subsection{Impact Chemistry and Bulk Atmospheric Composition}
\label{sec:impact-composition}

The effect of impacts is determined not simply by the frequency of impacts, but also by the initial atmospheric composition. As a consequence of the C/O ratio \citep{Rimmer2019b}, atmosphere composed only of CO$_2$ and N$_2$ will have very different chemical tracers of impact history than a planet with a more reducing atmosphere, typically in the form of CO/CH$_4$. It is important to note that the more reducing atmospheres result in greater yields of life's building blocks due to impacts \citep{Cleaves2008}.

We find that, as a result of the impact, about 72~\% of the methane is decomposed, while more stable gases such as carbon monoxide and nitrogen experience losses of 28~\% and 13~\% respectively. The initial and resulting compositions are shown in Table \ref{table:results1}. In previous studies referring exploration of this system by selected ion flow mass spectrometry, we estimated that the laser spark plasma produces also ppm amounts of ethanol, propadiene, propane, methanol, butadiene, propene, acetone and propanol \citep{Ferus2014,Ferus2009,Civis2016,Ferus2012,Civis2004,Civis2016}.

\subsection{Transmission Spectra of a Rocky Exoplanet experiencing Heavy Bombardment}
\label{sec:transmission}

\begin{figure}
\centering
\includegraphics[width=0.8\textwidth]{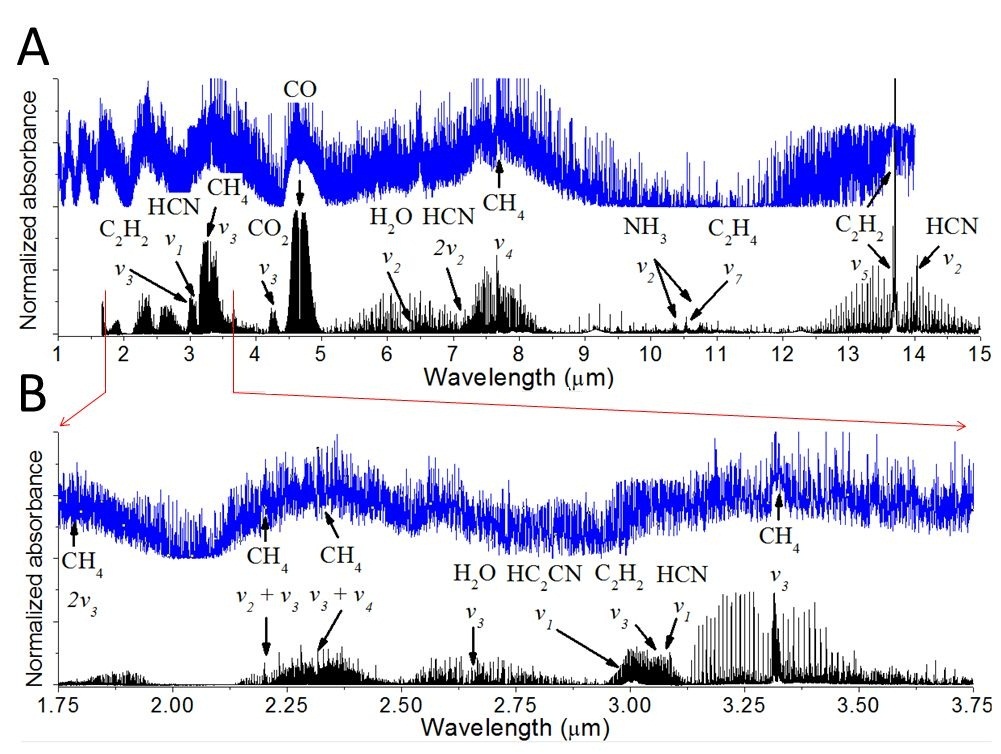}
\caption{Panel A shows infrared absorption spectra (from 1 -- 15 $\mu$m) of a gas phase CO + CH$_4$ + N$_2$ mixture exposed to an impact shock simulated by high-power laser PALS (black) and the transmission spectra for a temperature Earth-sized planet with a reducing atmosphere, experiencing Heavy Bombardment (blue). Panel B is the same from 1.75 -- 3.75 $\mu$m. The infrared absorption spectra were taken at room temperature. \label{fig:lab-spec}}
\end{figure}

\begin{figure}
\centering
\includegraphics[width=\linewidth]{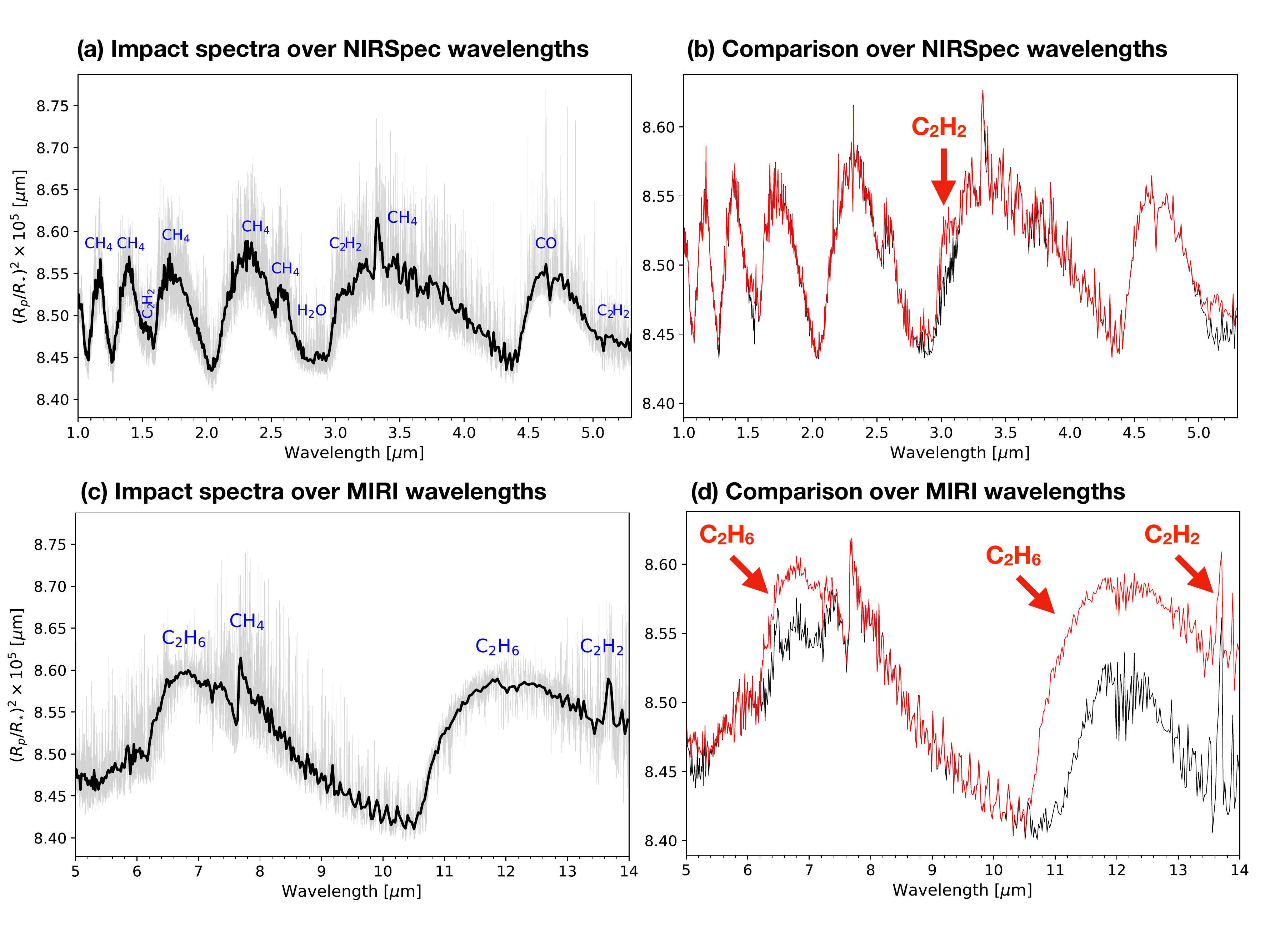}
\caption{Transmission spectra of a reducing (80\% \ce{N_2}, 10\% \ce{CO}, 10\% \ce{CH_4}) temperate atmosphere transformed by impacts (black), compared to the atmosphere without impacts (red) over the JWST NIRSpec (excluding optical) wavelength range (panels {\bf(a)} and {\bf(b)}), and MIRI wavelength range (panels {\bf(c)} and {\bf(d)}). The bulk atmospheric \ce{CH_4} covers all of the \ce{HCN} features and most \ce{C_2H_2} features. A prominent acetylene ``shoulder'' can be distinguished at 3.05 $\mu$m in transmission (top) and a prominent \ce{C_2H_2} feature can be seen at 13.5 $\mu$m (bottom). \ce{C_2H_6} signatures can also be detected, but are harder to distinguish between the atmosphere with and without impacts. \label{fig:transmission}}
\end{figure}

We use the chemical profiles (Fig. \ref{fig:photochem}) in a  TauREx radiative transfer model (see Methods), in order to predict transmission spectra for an Earth-sized planet around a Sun-like star (Fig's. \ref{fig:lab-spec},\ref{fig:transmission}). The first thing to note is that the large amounts of atmospheric methane obscure many of the other molecular features. Nevertheless, some of these features can be distinguished, even at a resolution of R=300. The bulk composition of the atmosphere can be detected: in addition to \ce{CH_4}, a strong \ce{CO} feature can be observed at 4.9 $\mu$m, \ce{H_2O} appears at 6.5 $\mu$m, and acetylene appears as a shoulder to the 3.3 $\mu$m \ce{CH_4} feature, and as a lone feature at 10.5 $\mu$m. All the \ce{HCN} features are masked by other features, and so we do not predict that \ce{HCN} is a good tracer for impacts on a reducing atmosphere. 

Fig. \ref{fig:transmission} emphasizes this acetylene shoulder, at roughly 3.05 $\mu$m, and this may be detectable using a combination of careful data reduction and atmospheric retrieval as with the nitrogen detection on HD 209458b \citep{MacDonald2017}. Hazes may obscure some features on this exoplanet, and thick hazes may render the 3.05 $\mu$m acetylene shoulder difficult to detect, but should not obscure the 10.5 $\mu$m feature \citep{Arney2016}. We predict detectable amounts of acetylene (\ce{C_2H_2}) as a tracer of impact-induced chemistry on methane-rich rocky exoplanets.


\section{Discussion}
\label{sec:discussion}

As the Sun formed from its molecular cloud, a disk of gas and dust formed around it. Over a few tens of millions of years the material in this disk coalesced to form the planets. This process occurred in several stages, eventually culminating in massive impacts on the proto-Earth. Upcoming observations of exoplanets afford a unique opportunity to observe directly consequences of heavy bombardment period during early stages of planetary evolution. Alongside other possible evidences of collisions in planetary systems \citep{Bonomo2019}, or maybe direct observation of large impact shock wave manifestations in exoplanetary atmospheres (emission spectra), detection of marker molecules produced in detectable quantities can be an alternative evidence of impact activity. These molecules can be identified only based on a combination of impact shock wave simulations with sophisticated models of planetary chemistry and simulation of corresponding transition spectra. Also, different atmospheric composition can provide different marker molecules. In the following chapters, we discuss identification of impacts markers within a reducing atmosphere.

\subsection{Analysis of the Atmospheric Chemistry after Impacts}
\label{sec:after-impact}

Here we provide a description of the atmospheric processing of the impact-produced species, HCN, \ce{NH_3} and \ce{C_2H_2}. We also discuss the formation of \ce{CH_2N_2}.

\begin{figure}
\centering
\includegraphics[width=\textwidth]{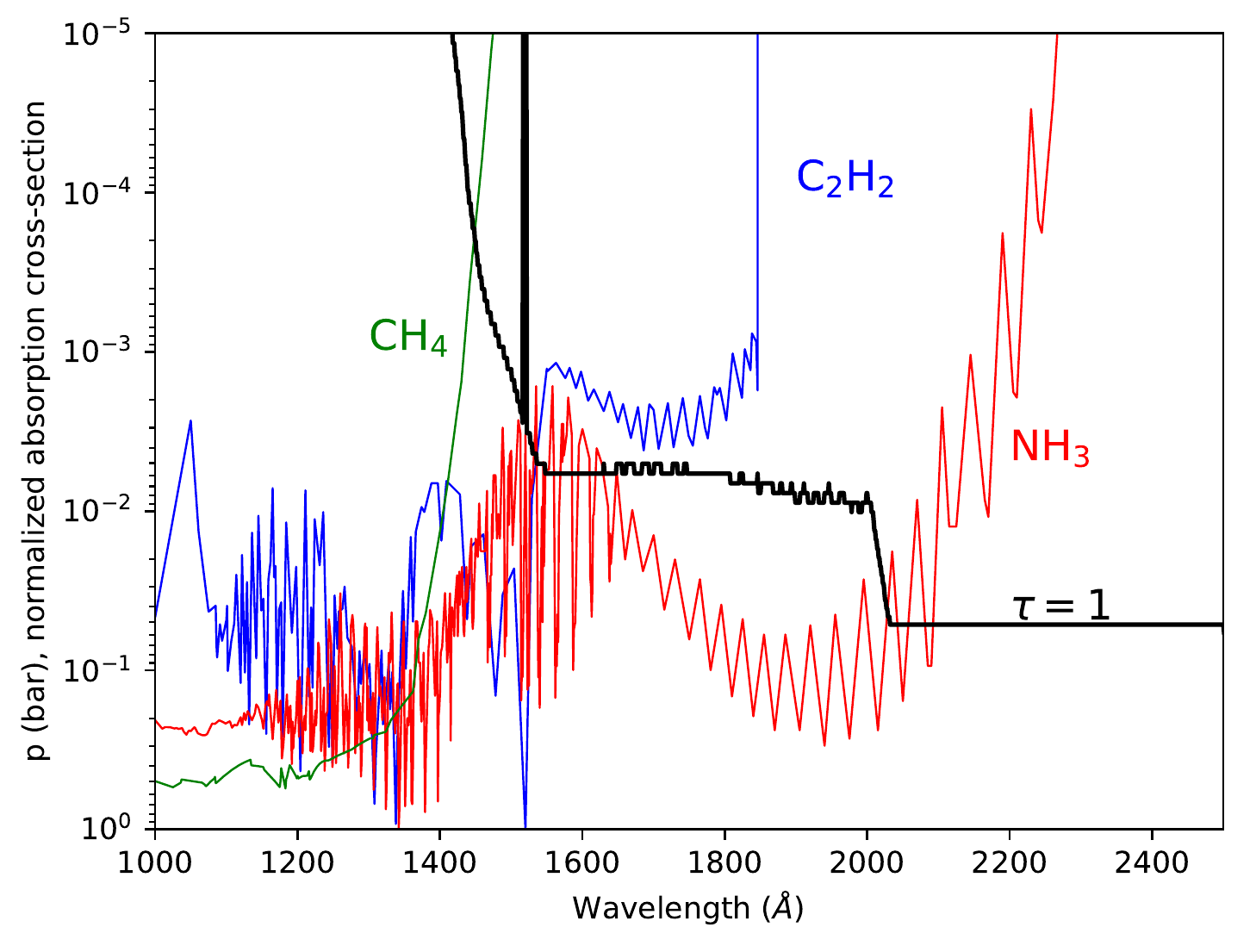}
\caption{A plot of the depth at which the optical depth, $\tau = 1$, as a function of pressure, as well as a plot of the normalized absorption cross-sections for \ce{CH_4}, \ce{NH_3} and \ce{C_2H_2} \label{fig:tauone}.}
\end{figure}

The lifetimes and profiles of all of these species are largely determined by atmospheric photochemistry, and the shape of their profiles is determined by how ultraviolet light at specific wavelengths penetrates the atmosphere. The depth at which photochemistry becomes important depends on the stellar spectrum, the bulk atmospheric composition, and the nature of the species in question. If there is an atmospheric window allowing near ultraviolet light to penetrate deep into the atmosphere, at wavelengths where another species will absorb these wavelengths, then photochemitry will be active deep within the atmosphere. This can be visualized by a $\tau = 1$ curve, showing where the optical depth, $\tau = 1$ as a function of pressure. We show this curve in Figure \ref{fig:tauone}.

\begin{figure}
\centering
\includegraphics[width=\textwidth]{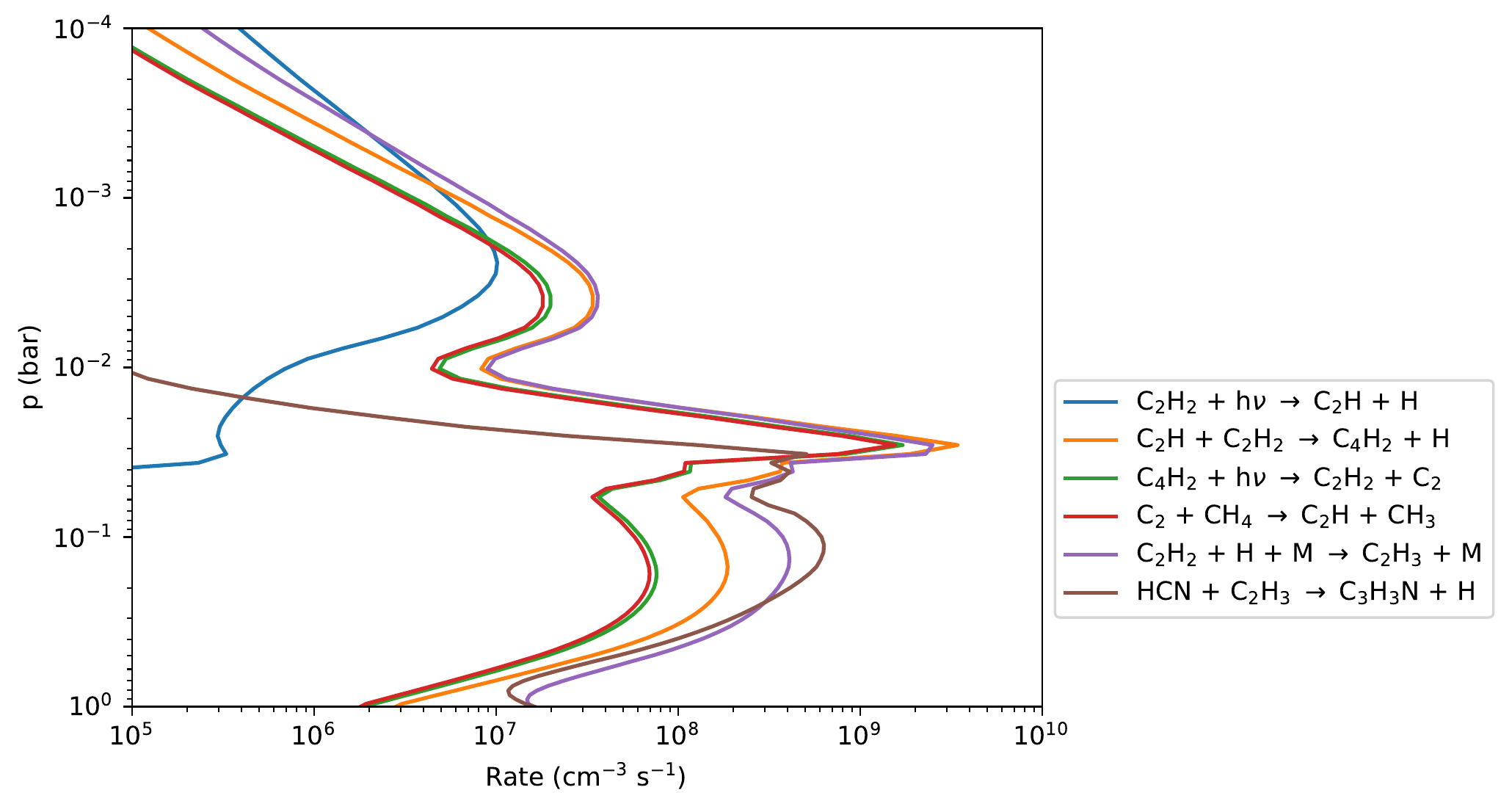}\\
\includegraphics[width=\textwidth]{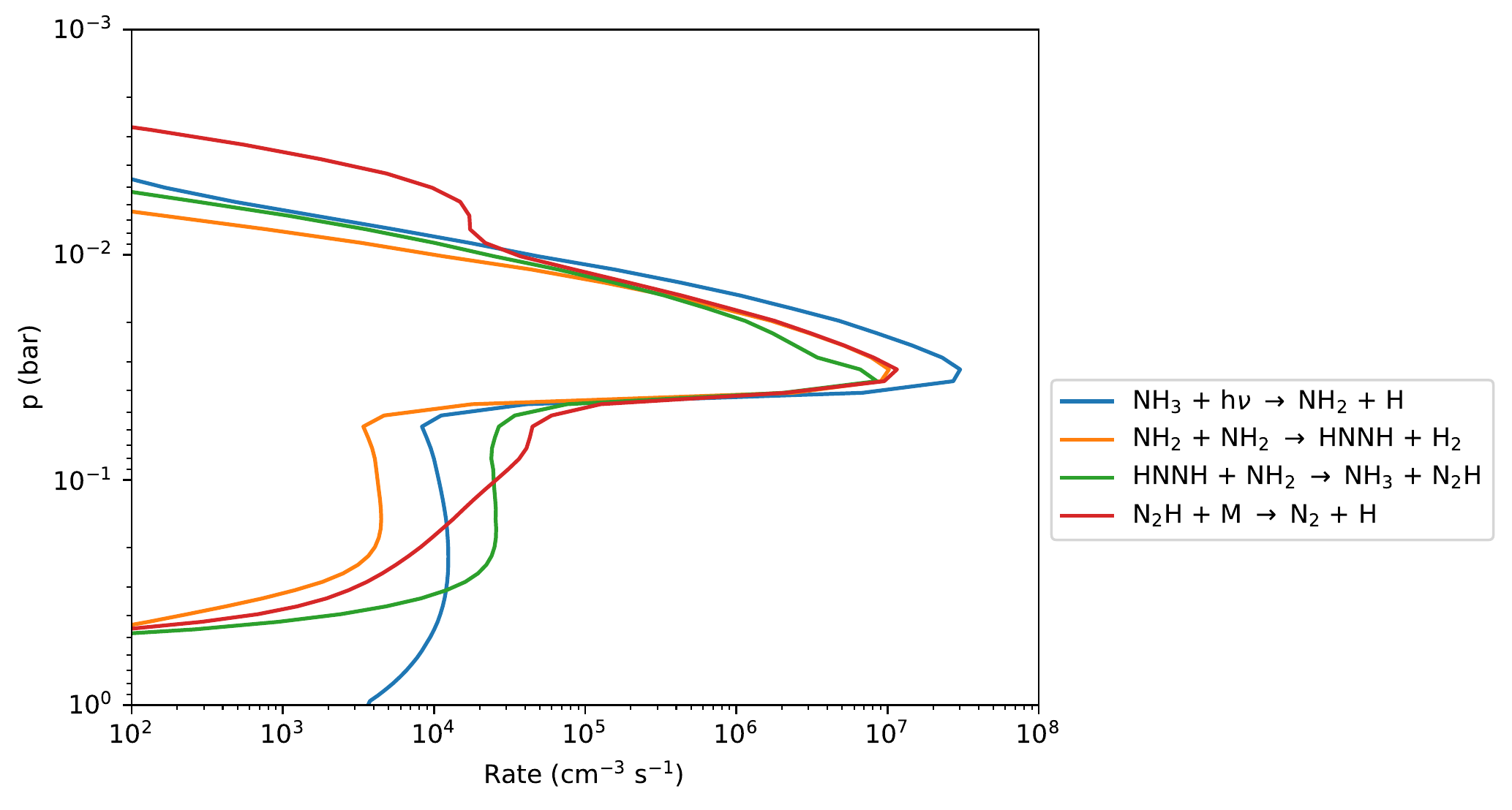}
\caption{Rates (cm$^{-3}$ s$^{-1}$) as a function of pressure (bar) for the dominant reactions that contribute to the destruction of \ce{HCN} and \ce{C_2H_2} (top figure) and \ce{NH_3} (bottom figure). \label{fig:rates}}
\end{figure}

At the concentrations produced via impacts, hydrogen cyanide and acetylene are relatively stable (over $>1$ year timescales) within a \ce{N_2}, CO, \ce{CH_4} dominated atmosphere. They are primarily destroyed photochemically. The rates for the dominant destructive pathways for the three primary impact-generated chemical species: acetylene, hydrogen cyanide and ammonia, are shown in Figure \ref{fig:rates}. The lifetimes of these species in the atmosphere are 250 years for acetylene, 30 years for hydrogen cyanide, and 15 years for ammonia. The ammonia lifetime is in general agreement with \citet{Kasting1982}. We have included these lifetimes in Table \ref{table:source}.

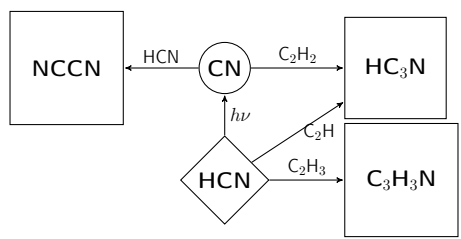
\begin{figure}
\centering
\resizebox{0.4\linewidth}{!}{
\begin{tikzpicture}[->,>=stealth',shorten >=1pt,auto,node distance=3cm,
                    thick,main node/.style={circle,draw,font=\sffamily\Large\bfseries},end node/.style={regular polygon,regular polygon sides=4,draw,font=\sffamily\Large\bfseries},impact node/.style={diamond,draw,font=\sffamily\Large\bfseries}]

  \node[main node] (2) {CN};
  \node[impact node] (3) [below of=2] {HCN};
  \node[end node] (4) [left=2cm of 2] {NCCN};
  \node[end node] (5) [right=2.5cm of 2] {HC$_3$N};
  \node[end node] (6) [right=2cm of 3] {C$_3$H$_3$N};

  \path[every node/.style={font=\sffamily\large}]
    (3) edge [above] node [right] {$h\nu$} (2)
    (2) edge [left] node [above] {HCN} (4)
    (2) edge [left] node [above] {C$_2$H$_2$} (5)
    (3) edge node [right] {C$_2$H} (5)
    (3) edge [left] node [above] {C$_2$H$_3$} (6);
\end{tikzpicture}}
\caption{Pathway to form cyanoacetylene, cyanogen and acrylonitrile from hydrogen cyanide. \label{fig:pathway-to-form-hcn}}
\end{figure}

We now discuss the chemical pathways that lead to the destruction of \ce{HCN}, \ce{NH_3} and \ce{C_2H_2}. Much of the \ce{HCN} is consumed by direct photodissociation or by reacting with products of acetylene, \ce{C_2H} and \ce{C_2H_3}, leading to the production of acrylonitrile (\ce{C_3H_3N}) and cyanoacetylene (\ce{HC_3N}), as shown in Fig. \ref{fig:pathway-to-form-hcn}.

The lifetimes of \ce{C_2H_2} and \ce{HCN} is largely regulated by \ce{C_2H_2} photodissociation:
\begin{equation}
\ce{C_2H_2} + h\nu \rightarrow \ce{C_2H} + \ce{H},
\end{equation}
followed by production of diacetylene via:
\begin{equation}
\ce{C_2H} + \ce{C_2H_2} \rightarrow \ce{C_4H_2} + \ce{H}.
\label{eqn:diacetylene}
\end{equation}
The diacetylene diffuses downward, and is photodissociated:
\begin{align}
\ce{C_4H_2} + h\nu \rightarrow& \ce{C_2} + \ce{C_2H_2},\\
& \ce{C_2H} + \ce{C_2H},
\end{align}
This destroys acetylene via Reaction (\ref{eqn:diacetylene}), although much of the acetylene is restored from the photochemical destruction of diacetylene, but the excess atomic hydrogen rapidly detroys acetylene via the reaction:
\begin{equation}
\ce{C_2H_2} + \ce{H} + \ce{M} \rightarrow \ce{C_2H_3},
\end{equation}
leading to \ce{C_2H_4} and \ce{C_2H_6}. In addition, diacetylene photochemical products do not always return to acetylene, but will react with other hydrocarbon fragments to form photochemical haze particles.

Ammonia is far less stable, and its reaction pathway is relatively simple. Ammonia photochemistry leads predominantly to the formation of molecular nitrogen:
\begin{align*}
3\Big(\ce{NH_3} + h\nu &\rightarrow \ce{NH_2} + \ce{H}\Big), \\
\ce{NH_2} + \ce{NH_2} &\rightarrow \ce{HNNH} + \ce{H_2}, \\
\ce{HNNH} + \ce{NH_2} &\rightarrow \ce{NH_3} + \ce{N_2H}, \\
\ce{N_2H} + \ce{M} &\rightarrow \ce{N_2} + \ce{H} + \ce{M}, \\
2\Big(\ce{H} + \ce{H} + \ce{M} &\rightarrow \ce{H_2} + \ce{M}\Big),\\
-----&-----\\
2\ce{NH_3} + 3h\nu &\rightarrow \ce{N_2} + 3\ce{H_2}.
\end{align*}
Much of the atomic hydrogen goes on to form \ce{H_2}, as shown in the scheme above, but a significant fraction reacts with \ce{CO} in a pathway to generate formaldehyde (see Fig. \ref{fig:pathway-to-form-nh3}) in addition to contributing to the hydrocarbon photochemistry.

We do, however, observe the efficient formation of ethane, given the quantity of available free hydrogen. Ethane is generated from acetylene in the upper atmosphere by the reaction scheme:
\begin{align*}
4 \big( \ce{CH_4} + h\nu &\rightarrow \ce{CH_3} + \ce{H}\big), \\
2 \big( \ce{C_2H_2} + \ce{H} + \ce{M} &\rightarrow \ce{C_2H_3} + \ce{M}, \\
2 \ce{C_2H_3} &\rightarrow \ce{C_2H_2} + \ce{C_2H_4}, \\
2 \big( \ce{C_2H_4} + \ce{H} + \ce{M} &\rightarrow \ce{C_2H_5} + \ce{M}, \\
2 \ce{C_2H_5} &\rightarrow \ce{C_2H_4} + \ce{C_2H_6}, \\
2 \big( 2 \ce{CH_3} + \ce{M} &\rightarrow \ce{C_2H_6} + \ce{M}, \\
-----&-----\\
\ce{C_2H_2} + 4\ce{CH_4} + 4h\nu &\rightarrow 3\ce{C_2H_6}
\end{align*}
The ethane is also relatively stable, but is photodissociated, and the methyl radicals can either react to reform ethane, can react with hydrogen to reform methane, or can go on to participate in the haze chemistry.

\begin{figure}
\centering
\resizebox{0.7\linewidth}{!}{
\begin{tikzpicture}[->,>=stealth',shorten >=1pt,auto,node distance=3cm,
                    thick,main node/.style={circle,draw,font=\sffamily\Large\bfseries},end node/.style={regular polygon,regular polygon sides=4,draw,font=\sffamily\Large\bfseries},impact node/.style={diamond,draw,font=\sffamily\Large\bfseries}]

  \node[main node] (1) {C$_2$H};
  \node[impact node] (2) [below of=1] {C$_2$H$_2$};
  \node[main node] (3) [below=2cm of 2] {C$_2$H$_3$O};
  \node[main node] (4) [right of=1] {C$_4$H$_2$};
  \node[end node] (5) [above of=4] {Haze};
  \node[main node] (6) [right of=2] {C$_2$H$_3$};
  \node[end node] (7) [right of=6] {C$_2$H$_4$};
  \node[end node] (8) [above right=0.5cm and -1.5cm of 7] {C$_3$H$_3$O};
  \node[main node] (9) [right of=7] {C$_2$H$_5$};
  \node[end node] (10) [right of=9] {C$_2$H$_6$};
  \node[main node] (12) [right=1.5cm of 3] {C$_2$H$_4$O};
  \node[main node] (13) [right=1.5cm of 12] {CH$_3$CH$_2$O};
  \node[end node] (14) [right=2cm of 13] {Acetone};

  \path[every node/.style={font=\sffamily\large}]
    (2) edge [above] node [right] {$h\nu$} (1)
    (1) edge [right] node [above] {C$_2$H$_2$} (4)
    (4) edge node [right] {Nucleation} (5)
    (2) edge node [right] {OH} (3)
    (2) edge [right] node [above] {H} (6)
    (6) edge [left] node [above] {H} (7)
    (6) edge node [left] {CO} (8)
    (7) edge node [above] {H} (9)
    (7) edge [bend right] node [below] {H$_2$} (10)
    (9) edge node [above] {H} (10)
    (3) edge node [above] {CH$_4$} (12)
    (12) edge node [above] {H} (13)
    (13) edge node [above] {H, H$_2$} (14);
\end{tikzpicture}}
\caption{Hydrocarbon chemistry, haze formation, and the atmospheric synthesis of acetone. \label{fig:pathway-to-form-c2h2}}
\end{figure}
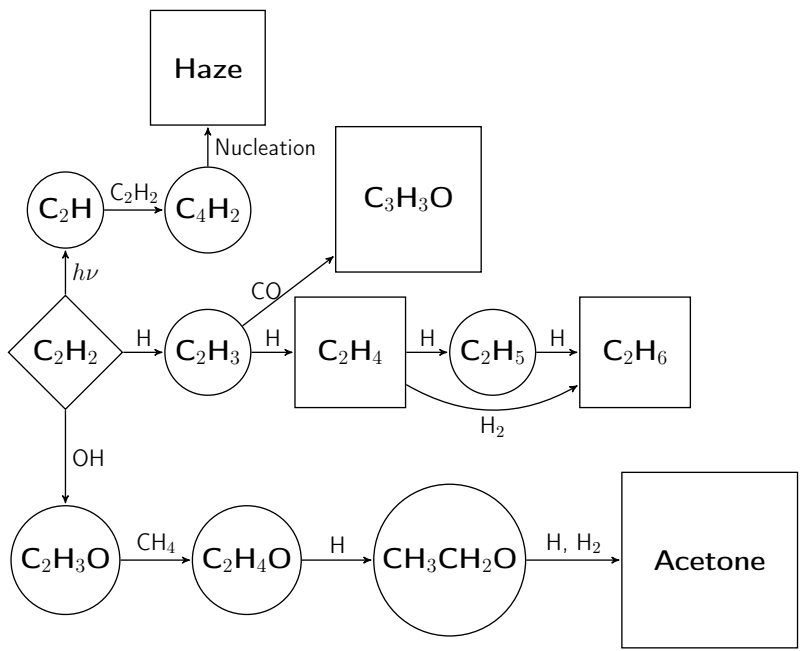

Alternatively, acetylene can be stepwise hydrated (attacked by a hydroxyl radical and then hydrogenated) to form acetone. This latter series of reactions is not efficient enough to produce observable quantities of acetone. Both of these pathways are shown in Fig. \ref{fig:pathway-to-form-c2h2}.

\begin{figure}
\centering
\resizebox{0.4\linewidth}{!}{
\begin{tikzpicture}[->,>=stealth',shorten >=1pt,auto,node distance=3cm,
                    thick,main node/.style={circle,draw,font=\sffamily\Large\bfseries},end node/.style={regular polygon,regular polygon sides=4,draw,font=\sffamily\Large\bfseries},impact node/.style={diamond,draw,font=\sffamily\Large\bfseries}]

  \node[impact node] (3) {NH$_3$};
  \node[main node] (5) [right of=3] {H};
  \node[main node] (6) [right=1cm of 5] {HCO};
  \node[end node] (7) [right=2.5cm of 6] {H$_2$CO};
  \node[end node] (8) [below of=6] {H$_2$};
  \node[end node] (9) [below of=7] {H$_2$};

  \path[every node/.style={font=\sffamily\large}]
    (3) edge [left] node [above] {$h\nu$} (5)
    (5) edge [right] node [above] {CO} (6)
    (6) edge [left] node [above] {HCO} (7)
    (6) edge node [left] {HCO} (8)
    (7) edge node [right] {$h\nu$} (9);
\end{tikzpicture}}
\caption{The enhancement of atomic hydrogen from ammonia photodissociation and its participation in the formation of formaldehyde. \label{fig:pathway-to-form-nh3}}
\end{figure}
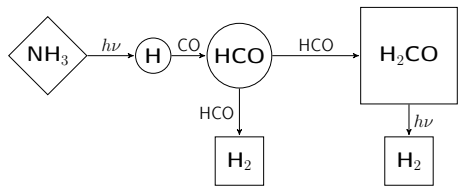

The pathway for forming diazomethane is straightforward:
\begin{align}
\ce{CH_4} + h\nu &\rightarrow \ce{^1CH_2} + \ce{H_2}, \\
\ce{^1CH_2} + \ce{N_2} + \ce{M} &\rightarrow \ce{CH_2N_2} + \ce{M}, \\
\ce{^1CH_2} + \ce{M} &\rightarrow \ce{^3CH_2} + \ce{M}, \\
\ce{^3CH_2} + \ce{N_2} + \ce{M} &\rightarrow \ce{CH_3N_2} + \ce{M},
\end{align}
where \ce{M} is any third body. The pathway for producing \ce{CH_3N_2} from singlet methane is given by \citet{Xu2010}, and the pathway from triplet methane by \citet{Braun1970}. We use the rate constants given by these references, for reactions which were already incorporated into STAND2015 \citep{Rimmer2016}, but as reverse reactions.

\subsection{The Predicted Effect of Bulk Atmospheric Chemistry on Impact Chemistry}
\label{sec:bulk-chemistry}

Rocky exoplanets may have a wide range of bulk atmospheric compositions, and it is important to explore how our results are expected to change for a range of atmospheres. Here we use chemical equilibrium as our guide. Although impact-generated chemistry does not reproduce equilibrium at any temperature, the results presented here and in \citet{Ferus2017} indicate that trends in the experimental results follow trends in chemical equilibrium at temperatures $\sim 2000 -- 5000$ K. These trends allow us to explore predicted results over a range of compositions prohibitively broad in terms of experimental time and cost. This is also an important exercise for contextualizing our experimental results within a hypothetical exoplanet atmosphere, which is much more rich in \ce{N_2}.

\begin{figure}
\centering
\includegraphics[width=0.45\textwidth]{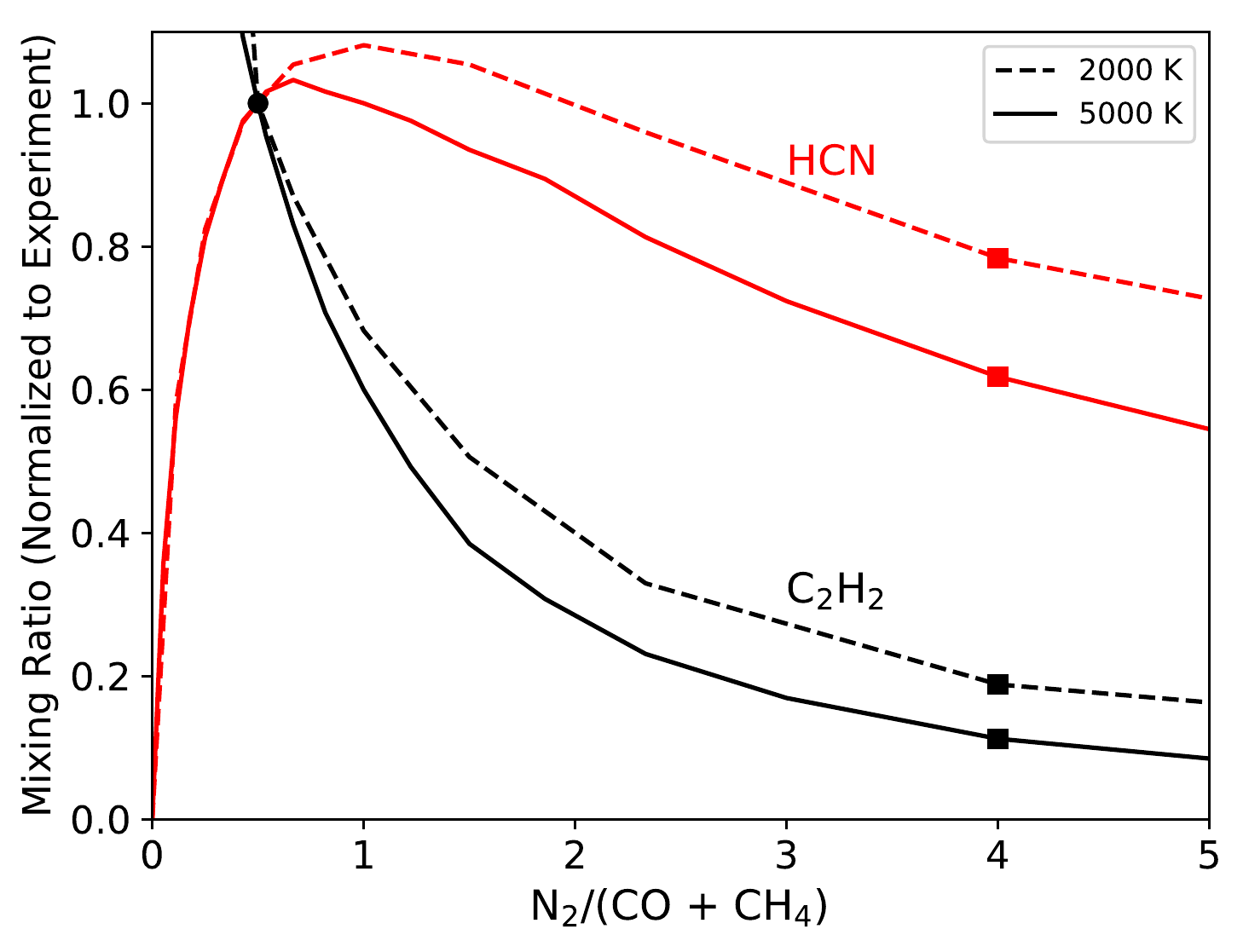}
\includegraphics[width=0.45\textwidth]{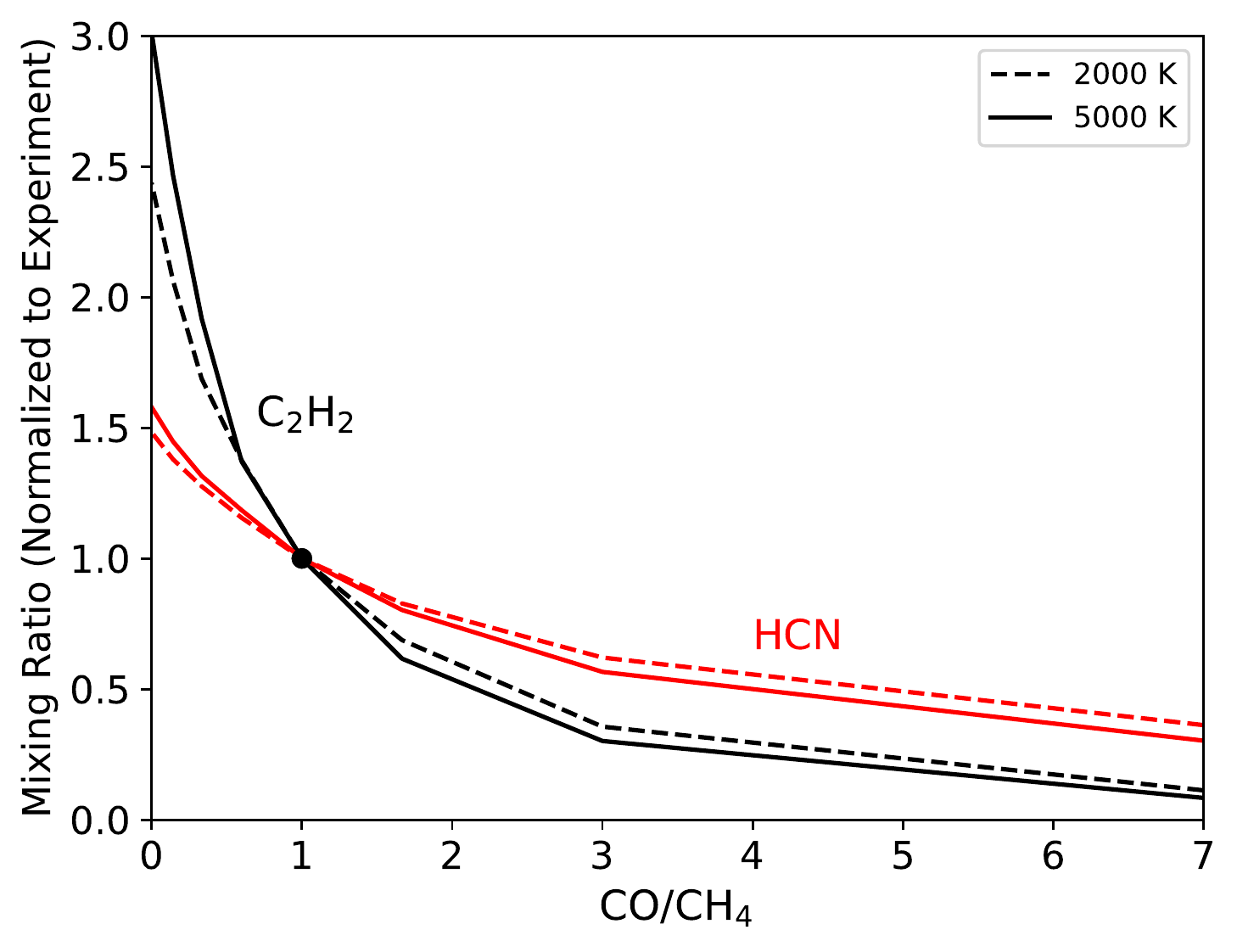}\\
\includegraphics[width=0.45\textwidth]{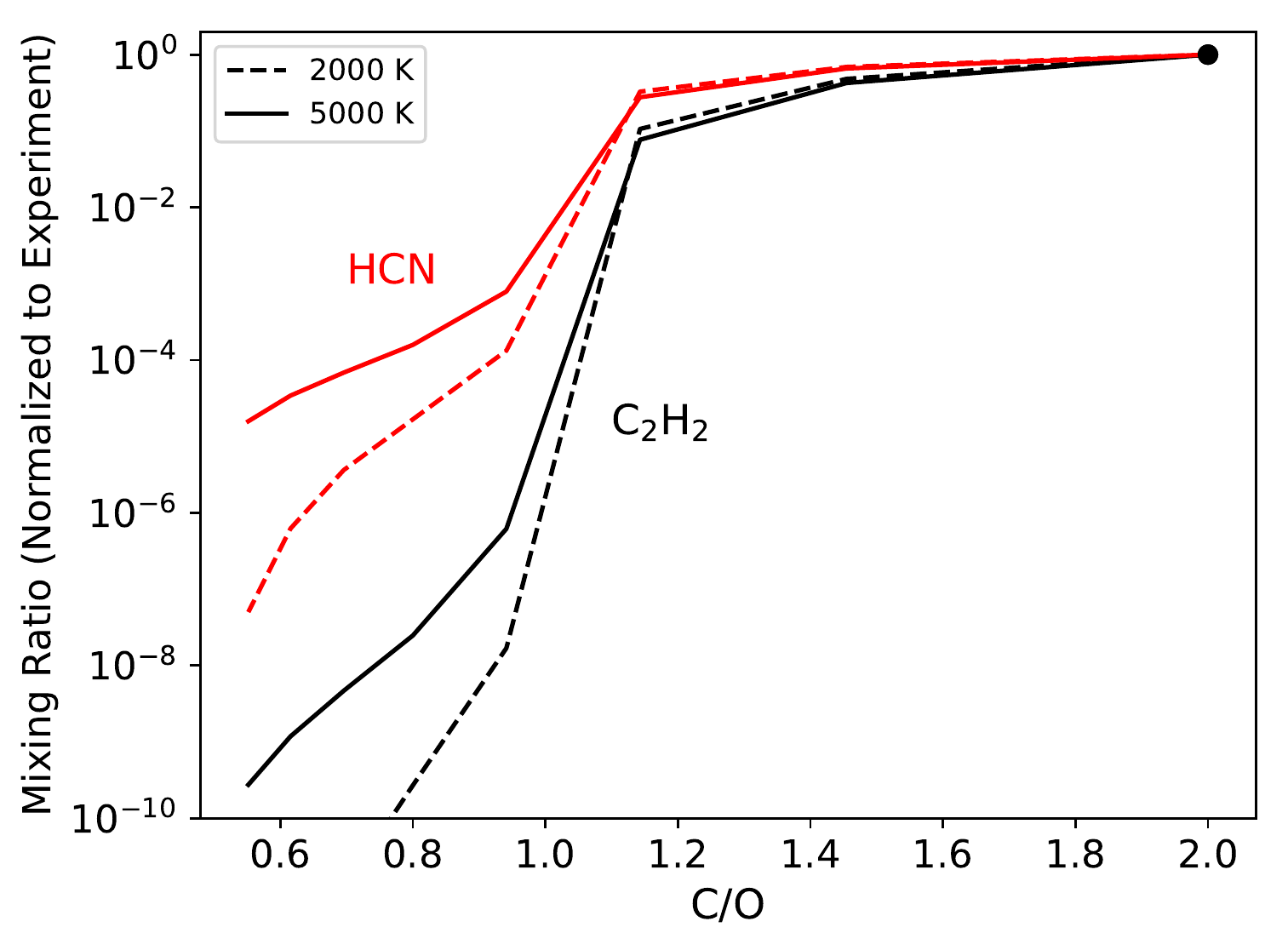}
\caption{Trends in thermochemical equilibrium mixing ratios of \ce{HCN} (red) and \ce{C_2H_2} (black), normalized to experimental values, as a function of the ratios \ce{N_2}/(\ce{CO} + \ce{CH_4}) (top left), \ce{CO}/\ce{CH_4} (top right), and \ce{C}/\ce{O} (bottom), for temperature of 2000 K (dashed) and 5000 K (solid). Points in all three figures indicate the experimental ratios, and squares indicate the ratios for our hypothetical exoplanet, only relevant for the left figure. The y-axis for the right figure is logarithmic.  \label{fig:var-bulk-chemistry}}
\end{figure}

We solve for chemical equilibrium at atmospheric surface pressure of 1 bar at 300 K and equilibrium temperatures of 2000 K (6.67 bar) and 5000 K (16.7 bar), varying the amount of \ce{CO}, \ce{CH_4}, \ce{N_2} and \ce{CO_2}, and express our results in terms of three different ratios:
\begin{itemize}
\item $\ce{N_2}/(\ce{CO} + \ce{CH_4})$, where $\ce{CO_2} = 0$, $\ce{CO} = \ce{CH_4}$, and \ce{N_2} mixing ratio is varied from 0 to 0.95.\\
\item $\ce{CO}/\ce{CH_4}$, where \ce{N_2} mixing ratio is fixed at 0.333, $\ce{CO_2} = 0$, and \ce{CO} and \ce{CH_4} mixing ratios are varied between 0 and 0.667.\\
\item $\ce{C}/\ce{O}$,  where \ce{N_2} mixing ratio is fixed at 0.333, $\ce{CO} = \ce{CH_4}$, and \ce{CO_2} mixing ratio is varied from 0 to 0.667.
\end{itemize}
The results of these calculations are plotted in Fig. \ref{fig:var-bulk-chemistry}. The results are not predicted to be very sensitive to changes in \ce{N_2}/(\ce{CO} + \ce{CH_4}) and \ce{CO}/\ce{CH_4}. The changes in abundance are within an order of
magnitude for both species over a wide range of values, including the
values selected for our hypothetical exoplanet. The \ce{C}/\ce{O} ratio, however, makess a big difference in the predicted results. We predict potentially observable quantities of acetylene (\ce{C_2H_2}) down to somewhere between $\ce{C}/\ce{O} \sim 1.1 - 1.5$. Below $\ce{C}/\ce{O} \sim 1.1$, \ce{C_2H_2} abundances drop off several orders of magnitude.

\subsection{Implications for Prebiotic Chemistry}
\label{sec:prebiotic}
Hydrogen cyanide is a necessary feedstock molecule for photochemical reaction pathways leading to the selective high-yield synthesis of pyrimidine nucleosides, amino acids and lipid precursors \citep{Ritson2012,Patel2015}. High rates of impacts provide a challenge for this synthetic pathway, due to the thick hazes that will result, which risk obscuring the 200-280 nm light needed for this chemistry. It has not been determined how effective these hazes will be in shielding the surface from the UV light. Some prebiotic chemistry has been shown to occur within the haze particles themselves \citep{Horst2012}. Additionally, HCN can be stored on the surface in the form of ferrous cyanide and other organometallic complexes, to be liberated by liquid water once the haze clears, after which the HCN photochemistry can proceed unimpeded \citep{Ritson2018}. Finally, the HCN produced, whether in its original form, as a salt, or as ferrous cyanide, can be subducted and would later be outgassed from vents again as HCN and other products, after the haze has cleared \citep{Rimmer2019a}.

\subsection{Conclusions}
\label{sec:conclusion}

The generation of \ce{C_2H_2}, HCN and other products takes place in an extremely methane-rich atmosphere, and it is an open question about how this amount of methane could be sustained through an era of high impacts. Experiments suggest that the methane is destroyed by impacts, although not efficiently enough to significantly change its surface abundance. Coupled with UV light, energetic particles and subsequent loss through formation of haze particles and escape of \ce{H_2}, it seems that a significant amount of methane must be outgassed at this time to restore that which was lost. Investigations into the redox state of the Hadean crust \citep{Yang2014}, and observations of Titan (see Table \ref{table:results1}) and preliminary spectra from 55 Cnc e \citep{Tsiaras2016}, suggests this possibility can be realized both on the Early Earth and now on other moons and planets. 

Comparing spectra with and without impact shows that photochemistry alone is not sufficient to produce detectable quantities of \ce{C_2H_2}, as is apparent from Figure \ref{fig:transmission}. This does not mean the presence of \ce{C_2H_2} is an unambiguous indicator of impacts. There are local scenarios where \ce{C_2H_2} could be outgassed in large quantities from extremely reducing volcanic plumes, similar to those of mud volcanos on Earth \citep{Rimmer2019a}. If a planet's magmatic and crustal chemistry were very different from Earth's, it may be that these local volcanic plumes would be ubiquitous, in which case \ce{C_2H_2} could be one of the dominant atmospeheric species. Acetylene, therefore, is a reliable impact signature only when the planet-star system is taken in context.

\acknowledgments
This paper has been published as part of research series supported by the Czech Science Foundation within the project reg. no. 19-03314S and ERDF/ESF "Centre of Advanced Applied Sciences" (No. CZ.02.1.01/0.0/0.0/16\_019/0000778. Our thanks go to Ji\v{r}\'{i} Sk\'{a}la, Miroslav Pfeifer, Pavel Prchal and Jakub Mare\v{s} for a valuable technical assistance at the PALS facility. The work at the PALS facility was financially supported by the Czech Ministry of Education (Grants LTT17015, CZ.02.1.01/0.0/0.0/16\_013/0001552 and LM2015083). Part of this project was also financially supported by project GAUK 1674218.  P.~B.~R. thanks the Simons Foundation for support under SCOL awards 59963. S. N. Y. and J. T. thanks the support of the UK Science and Technology Research Council (STFC) No. ST/M001334/1 and ST/R000476/1. This project has received funding from the European Research Council (ERC) under the European Union Horizon 2020 research and innovation programme (grant agreement No 758892, ExoAI). ATA thanks NERC for funding through the National Centre for Atmospheric Science.

\end{document}